\documentclass[conference]{IEEEtran}
\IEEEoverridecommandlockouts
\usepackage{cite}
\usepackage{amsmath,amssymb,amsfonts}
\usepackage[ruled,lined,linesnumbered,noend]{algorithm2e}
\SetKw{KwTo}{to}
\SetKw{KwBy}{by}
\usepackage{graphicx}
\usepackage{textcomp}
\usepackage{xcolor}
\usepackage{listings}
\usepackage{pifont}
\usepackage{booktabs}
\usepackage{multirow}
\usepackage[hidelinks]{hyperref}
\usepackage{caption}
\usepackage{enumitem}
\usepackage{orcidlink}
\usepackage{fancyhdr}

\def\BibTeX{{\rm B\kern-.05em{\sc i\kern-.025em b}\kern-.08em
    T\kern-.1667em\lower.7ex\hbox{E}\kern-.125emX}}

\definecolor{darkgreen}{RGB}{0,127,0}
\definecolor{darkred}{RGB}{127,0,0}
\definecolor{shadecolor}{RGB}{250, 250, 250}
\begin{document}

\title{
EmuGEMM: Fused Tensor Core Kernels for Precision Emulation in Matrix Multiplication
}

\author{
\IEEEauthorblockN{Denghui Lu\,\orcidlink{0000-0003-0977-3635},
Alexander Maeder\,\orcidlink{0009-0003-4420-5593},
Mathieu Luisier\,\orcidlink{0000-0002-2212-7972},
Alexandros Nikolaos Ziogas\,\orcidlink{0000-0002-4328-9751}}
\IEEEauthorblockA{\textit{D-ITET, ETH Zurich}, Zurich, Switzerland \\
\{denghuilu, almaeder, mluisier, alziogas\}@iis.ee.ethz.ch}
}



\maketitle

\thispagestyle{fancy}
\lhead{}
\rhead{}
\chead{}
\lfoot{\footnotesize{SC26, November 15-20, 2026, Chicago, Illinois, USA.
\copyright~2026 IEEE. Personal use of this material is permitted.  Permission from IEEE must be obtained for all other uses, in any current or future media, including reprinting/republishing this material for advertising or promotional purposes, creating new collective works, for resale or redistribution to servers or lists, or reuse of any copyrighted component of this work in other works.}}
\rfoot{}
\cfoot{}
\renewcommand{\headrulewidth}{0pt}
\renewcommand{\footrulewidth}{0pt}

\begin{abstract}
Modern GPUs devote an increasing silicon budget to low-precision matrix-multiplication units, widening the precision--throughput gap for scientific computing workloads.
Ozaki Schemes I and II offer an alternative by reconstructing high-precision general matrix multiplication (GEMM) from low-precision operations, yet existing implementations leave substantial performance untapped.
In particular, intermediate results are repeatedly materialized in global memory, making data movement the dominant bottleneck.
We present \textit{EmuGEMM}, fused integer Tensor Core kernels for NVIDIA Hopper and Blackwell GPUs that eliminate redundant memory round-trips in both Ozaki schemes.
Using Scheme~I, EmuGEMM sustains up to 1{,}639~Top/s on Hopper (83\% of INT8 peak) and 3{,}654~Top/s on Blackwell (81\%).
For large matrices, EmuGEMM surpasses cuBLAS TF32 throughput by up to $1.4\times$ on Hopper and $1.7\times$ on Blackwell, at comparable accuracy.
Using Scheme~II, EmuGEMM extends to complex arithmetic and outperforms cuBLAS ZGEMM by up to $2.3\times$ on Hopper and $5.5\times$ on Blackwell.

\end{abstract}


\section{Introduction}
\label{sec:intro}

\begin{figure}[!t]
    \centering
    \includegraphics[width=\columnwidth]{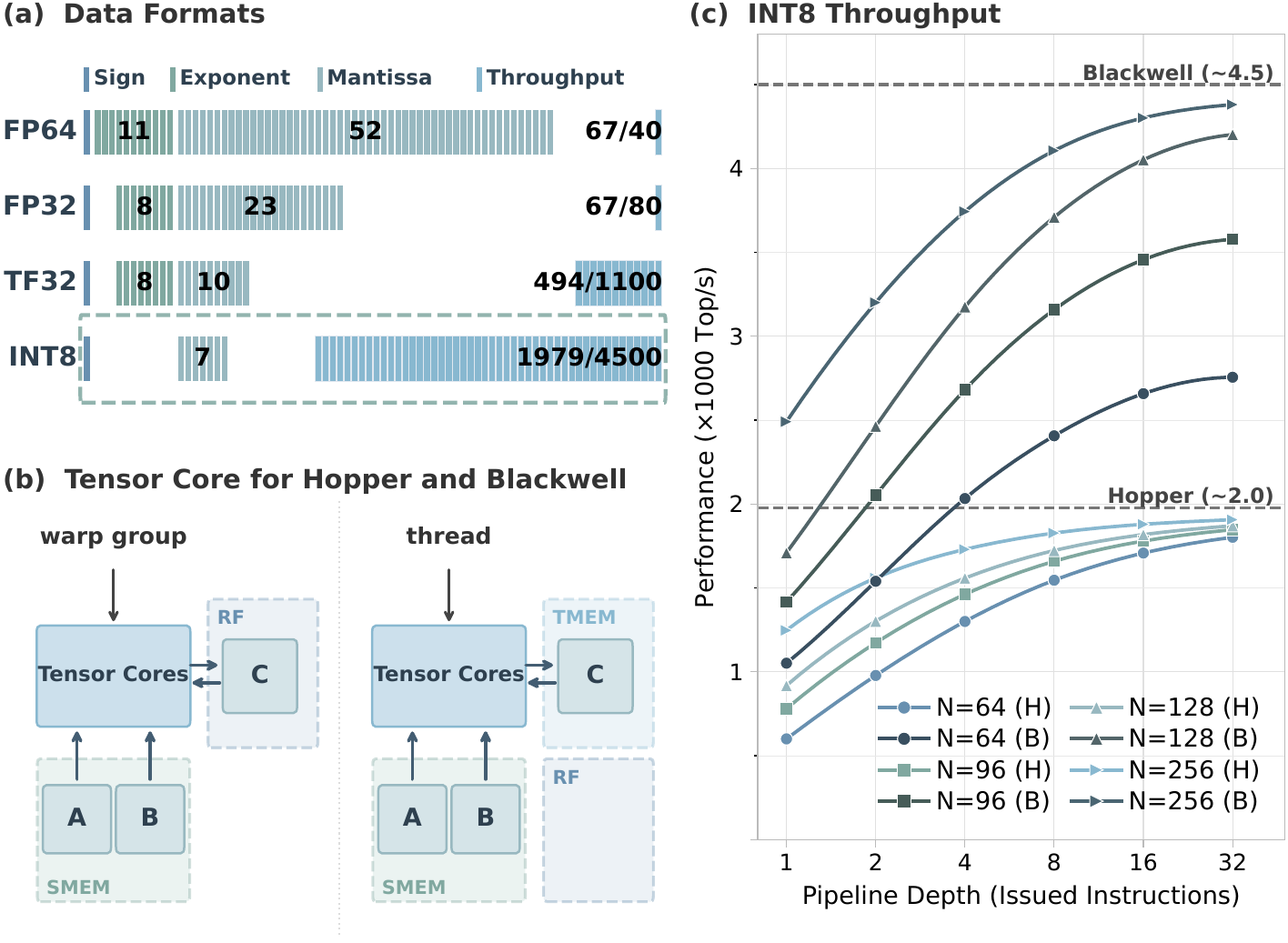}
    \caption{\textbf{GPU data formats and TC characterization.}
    (a)~Bit layout and peak throughput (Top/s, NVIDIA GH200\,/\,B200) for four arithmetic formats; INT8 delivers up to $112\times$ the throughput of FP64.
    (b)~TC instruction interface on Hopper (left) and Blackwell (right).
    (c)~Sustained INT8 throughput as a function of MMA pipeline depth~$\omega$ and tile width~$t_N$; dashed lines mark the hardware peak (${\sim}1{,}979$~Top/s on GH200, ${\sim}4{,}500$~Top/s on B200).}
    \label{fig:background}
    \vspace{-1.2em}
\end{figure}

General matrix multiplication (GEMM) is the computational backbone of modern high-performance computing (HPC) and machine learning (ML) workloads~\cite{vanzee-2015, jouppi-2017, dao2022flashattention, vetsch2025ab}.
The exponential growth of ML applications, in particular large language models (LLMs), has driven GPU manufacturers to dedicate an increasing silicon area to specialized low-precision matrix-multiplication units (MMUs), such as NVIDIA Tensor Cores (TCs)~\cite{nv_tensor_core} or AMD Matrix Cores~\cite{amd_cdna}.
Each of these architectures is capable of delivering an order-of-magnitude higher throughput on reduced-precision operands (e.g., FP8, INT8) than on double-precision floating-point (FP64)~\cite{nvidia_hopper}.
Notably, this trend is accelerating.
NVIDIA's Blackwell chips further increase the INT8-to-FP64 throughput ratio to 112$\times$ (Fig.~\ref{fig:background}(a)) from 30$\times$ on Hopper, corresponding to a near doubling of the gap within a single generation, pushed by the economic incentive to optimize GPUs for LLM workloads. The latter indeed significantly outweigh the demand for high-precision arithmetic.
As a consequence of this trajectory, scientific computing workloads requiring FP32 or even FP64 precision face growing challenges.
Domains such as computational chemistry or materials, climate modeling, finite-element analysis, and numerical linear algebra cannot directly migrate to lower precision without sacrificing solution accuracy~\cite{higham2019simulating, blanchard2020mixed, abdelfattah-2021}.
For these communities, the sustained shift toward low-precision MMUs risks limiting their workloads to an increasingly small fraction of available hardware throughput.

A promising approach to bridging this precision--throughput gap is \emph{emulation}.
For example, the Ozaki Scheme~I~\cite{ozaki-1, ootomo-2024} splits floating-point matrices into multiple low-precision slices, so that each pairwise slice product is computed exactly by the MMUs.
The full-precision result is then reconstructed from these exact partial products.
Recently, Ozaki et al.~\cite{ozaki-2} proposed a complementary strategy that leverages the Chinese Remainder Theorem (CRT) to perform emulation via integer modular arithmetic rather than mantissa splitting, offering finer control over the number of decomposed GEMMs and thus the attainable precision.
This method is referred to as the Ozaki Scheme~II. Uchino et al.~\cite{uchino-2025c} further extended Scheme~II to complex arithmetic via the 3M method, which expresses all complex multiplications as three real-valued multiplications, each carried out through the CRT-based emulation.

Existing implementations of these emulation schemes~\cite{mukunoki-2020, ootomo-2024, ozaki-2, uchino-2025a} are typically far from optimal, ignoring available performance boosting strategies:
The overhead of multiple GEMM-kernel launches, repeated materialization of intermediate results in global memory, and suboptimal data layouts prevent the underlying hardware from being fully exploited.
The recent integration of Ozaki Scheme~I into cuBLAS~\cite{schwarz2026guaranteed} has led to impressive performance improvements, providing $2.3\times$ speedup over native FP64 GEMM routines on Blackwell B200.
On Hopper, however, cuBLAS emulation outperforms native FP64 TCs only at low slice counts (i.e., $p \leq 5$). Even in its most favorable configuration, the emulated FP64 throughput remains well below what the INT8 TC peak theoretically permits. Furthermore, cuBLAS is currently restricted to the Scheme~I, leaving Scheme~II unexplored.

In this paper, we present \textbf{EmuGEMM}, a high-performance emulated GEMM library that systematically addresses these limitations through hardware-aware kernel fusion and outperforms FP64 native GEMM on both the NVIDIA Hopper and Blackwell architectures. The key concept at EmuGEMM's core is straightforward. In Scheme~I with $p$ slices, the emulation requires $p(p+1)/2$ independent INT8 GEMMs, as discussed in Section~\ref{subsubsec:back-ozaki-i}.
When launched as separate kernels, they reload the operand slices from global memory, resulting in $\mathcal{O}(p^2)$ total slice loads.
Fusing all products into a single kernel allows each slice to be loaded only once, reducing the traffic to $\mathcal{O}(p)$.
Scheme~II suffers from a similar bottleneck: 
Each of the $p$ independent INT8 GEMMs produces an INT32 result that must then undergo a modular reduction to extract its INT8 residue (Section~\ref{sec:back-ozaki-ii}).
When launched as separate kernels, each INT32 output incurs a full global memory round-trip, tripled for complex GEMM under 3M.
Fusing the reduction into each GEMM kernel keeps these intermediates on chip.

In both Schemes~I and II, kernel fusion is not merely an optimization, but a prerequisite to fully realize the theoretical throughput advantage of emulation.
Accordingly, EmuGEMM comprises two fused kernel implementations, each designed according to the target hardware specifications. With this respect, we make the following contributions:

\begin{itemize}[leftmargin=*]
\item \textbf{Interleaved data layout.}
  We co-design the decomposed slices' global memory layout to align with the MMU's tile dimensions without extra data movement, allowing execution of all $p(p{+}1)/2$ slice-pair products within a single kernel.
  This layout is format-agnostic and applies to any Ozaki-style decomposition.
\item \textbf{EmuGEMM-I: fused Scheme~I kernel.}
  Building on this layout, we implement a persistent kernel for NVIDIA Hopper and Blackwell GPUs, executing all $p(p{+}1)/2$ products via a triangular MMA schedule over $p$ on-chip accumulators, followed by an in-register shift-reduce epilogue that reconstructs the output without further global memory traffic.
\item \textbf{EmuGEMM-II: fused Scheme~II kernel.}
  We fuse the INT32-to-INT8 modular reduction directly into the GEMM on integer TCs, eliminating per-modulus global memory round-trips.
  Expanding on this, we implement CRT-based 3M complex multiplication as a fused GEMM on integer TCs, with error-free outputs during the multiplication stage.
\end{itemize}

Leveraging microarchitectural features of NVIDIA Hopper and Blackwell GPUs for both kernels, EmuGEMM-I surpasses cuBLAS TF32 throughput by $1.4\times$ on Hopper and $1.7\times$ on Blackwell at comparable accuracy.
Our kernels sustain up to 1{,}639~Top/s of INT8 throughput on Hopper (83\% of peak) and 3{,}654~Top/s on the B200 (81\%), exceeding cuBLAS native INT8 GEMM by $1.1\times$ and far surpassing cuBLAS Scheme~I emulation by $1.4\times$. 
For complex arithmetic, EmuGEMM-II outperforms cuBLAS native ZGEMM by up to $2.3\times$ on Hopper and $5.5\times$ on Blackwell with the 3M algorithm.

The rest of this paper is organized as follows.
Section~\ref{sec:back} reviews the relevant TC microarchitectures and the Ozaki schemes.
The interleaved data layout and the two fused-kernel designs are presented in Section~\ref{sec:emu-i} and Section~\ref{sec:emu-ii}.
Section~\ref{sec:evaluation} evaluates \textit{EmuGEMM} on GH200 and B200 GPUs and compares these results against cuBLAS native and emulated GEMM, and discusses current limitations.
We close with related work in Section~\ref{sec:related} before concluding in Section~\ref{sec:conclusions}.

\section{Background}
\label{sec:back}

This section presents the hardware and algorithmic foundations of our kernel designs, starting with 
the TC architecture (Section~\ref{sec:back-gpu}), then the tiled GEMM execution model that defines the resource budget (Section~\ref{sec:back-gemm}), and finally the Ozaki decomposition schemes that determine the computational structure (Section~\ref{sec:back-ozaki}).

\subsection{GPU Architecture and Tensor Cores}
\label{sec:back-gpu}

We target NVIDIA Hopper (SM90) and Blackwell (SM100) microarchitectures.
Each GPU comprises multiple \emph{streaming multiprocessors} (SMs). An SM executes one or more \emph{cooperative thread arrays} (CTAs, or thread blocks), within which threads can be partitioned into \emph{warpgroups} (WG).
Each SM contains \emph{Tensor Core} (TC) units that compute matrix products, along with on-chip \emph{shared memory} (SMEM) for staging input and output data and a \emph{register file} (RF) for holding thread-local state and intermediate values.
Blackwell SMs additionally provide \emph{tensor memory} (TMEM), a dedicated on-chip storage for accumulator data, separate from both RF and SMEM.
In the kernels considered in this paper, operand data reside in off-chip \emph{global memory} (GMEM) and are transferred to SMEM via \emph{tensor memory accelerator} (TMA) instructions, which perform asynchronous bulk copies between GMEM and SMEM without staging through registers.

On both GH200 and B200 GPUs, arithmetic throughput spans two orders of magnitude across data formats (Fig.~\ref{fig:background}(a)).
Among them, INT8 is especially attractive for precision emulation: An INT8$\times$INT8 matrix multiplication accumulated into INT32 is an exact integer computation without rounding error.
INT8 TCs are accessed via architecture-specific \emph{matrix-multiply-accumulate} (MMA) instructions (Fig.~\ref{fig:background}(b)). A single MMA multiplies a tile of size $t_M \times t_K$ by one of size $t_K \times t_N$ and accumulates the result into a $t_M \times t_N$ INT32 output tile.
On Hopper, we use \texttt{wgmma.mma\_async}, a WG-level instruction with $t_M{=}64$, $t_K{=}32$, and configurable $t_N$. In our kernels, operand tiles are staged in SMEM and the accumulator is held in RF.
On Blackwell, we use \texttt{tcgen05.mma}, with $t_M{=}128$, $t_K{=}32$, and configurable $t_N$; Operand tiles are again staged in SMEM, while the accumulator is held in TMEM.

\subsection{Tiled GEMM Kernel Model}
\label{sec:back-gemm}

\begin{table}[!t]
\caption{Tiled INT8 GEMM kernel model: notation, tile dimensions,
         and per-CTA resource budgets (in bytes).}
\label{tab:gemm-model}
\centering
\resizebox{\columnwidth}{!}{%
\begin{tabular}{llll}
\toprule
    Category & Symbol & Description & Value / Formula \\
\midrule
\multicolumn{4}{l}{\textcolor{gray}{\textit{%
    MMA instruction tile (architecture-dependent)}}} \\
\textbf{MMA}
    & $t_M$ & $M$-dimension & 64\,(H)\,/\,128\,(B) \\
    & $t_N$ & $N$-dimension & configurable \\
    & $t_K$ & $K$-dimension & 32 \\
\midrule
\multicolumn{4}{l}{\textcolor{gray}{\textit{%
    Kernel design choices}}} \\
\textbf{Tuning}
    & $\alpha$ & $M$-stacking factor
               & configurable \\
    & $\sigma$  & SMEM swizzle mode 
               & 1\,/\,2\,/\,4 \\
    & $\gamma$ & SMEM buffering stages 
               & configurable \\
\midrule
\multicolumn{4}{l}{\textcolor{gray}{\textit{%
    Derived CTA tile dimensions}}} \\
\textbf{CTA Tile}
    & $b_M$ & CTA $M$-dimension & $\alpha \, t_M$ \\
    & $b_N$ & CTA $N$-dimension & $t_N$ \\
    & $b_K$ & CTA $K$-dimension & $\sigma \, t_K$ \\
\midrule
\multicolumn{4}{l}{\textcolor{gray}{\textit{%
    Per-CTA resource consumption (bytes)}}} \\
\textbf{Budget}
    & $\mathrm{Acc}$ & Accumulator buffer
    & $4\,\alpha\,t_M\,t_N$ \\
    & $S_{\mathrm{op}}$ & Operand buffer in SMEM
    & $(b_M{+}b_N)\,b_K\,\gamma$ \\
    & $S_{\mathrm{epi}}$ & Epilogue buffer in SMEM
    & $4\,b_M\,b_N$ \\
\bottomrule
\end{tabular}%
}
\vspace{-1.5em}
\end{table}

High-performance GEMM kernels build on these TC instructions (Section~\ref{sec:back-gpu}) to compute $C = AB$ for matrices $A \in \mathbb{R}^{M \times K}$ and $B \in \mathbb{R}^{K \times N}$. Each CTA computes an output tile and iterates along the shared dimension~$K$ in steps of~$b_K$.
At each step, TMA loads the required tiles of the input matrix $A$ and $B$ from GMEM into SMEM accordingly, the CTA then issues multiple MMA instructions over the loaded data, and the results accumulate on-chip in RF (Hopper) or TMEM (Blackwell). 
After the $K$-loop completes, an epilogue stage converts the accumulator to the desired output format and writes back to GMEM.
To overlap TMA latency with computation, $\gamma$~successive operand tiles are simultaneously buffered in SMEM.
Table~\ref{tab:gemm-model} summarizes the design parameters and their implications for resources.
Each $K$-step issues $\omega = \alpha \cdot \sigma$ MMA instructions.
Sustained TC throughput is governed by two parameters: the number of MMA instructions issued per $K$-step, and the tile width~$t_N$. 
As Fig.~\ref{fig:background}(c) shows, throughput saturates once the instruction depth $\omega$ reaches~16 and degrades sharply below this threshold, particularly for narrow tiles~$t_N$. 
Maximizing~$\omega$ demands large tiles, but the INT32 accumulator ($\mathrm{Acc}$) and the multi-stage operand buffer ($S_{\mathrm{op}}$) compete for finite on-chip storage, bounding the tile size per SM.

\subsection{Precision Emulation via Matrix Decomposition}
\label{sec:back-ozaki}

The exact INT32 accumulation of INT8 TCs (Section~\ref{sec:back-gpu}) makes it possible to emulate high-precision arithmetic using low-precision hardware.
Ozaki et al.~\cite{ozaki-1, ozaki-2} exploit this property by decomposing each operand into INT8 slices, multiplying all relevant slice pairs via INT8 GEMMs, and reconstructing the full-precision result from the INT32 partial products.
Both Ozaki schemes implement this strategy, but differ in the decomposition method and, critically, in the number of INT8 GEMMs required.
This count determines the volume of intermediate results that must be managed on-chip or written to memory, and shapes the kernel fusion strategies.

\subsubsection{\textbf{Scheme~I}}
\label{subsubsec:back-ozaki-i}

Scheme~I decomposes each input operand by extracting consecutive $\beta$-bit segments ($\beta \leq 8$) from the mantissa, so that each segment fits into INT8 format. The original signed-slice method in~\cite{ootomo-2024} supports $\beta \leq 7$, while the unsigned-slice extension in~\cite{schwarz2026guaranteed} raises this limit to $\beta \leq 8$. Thus, every pairwise segment product is exact in INT32, and the positional weights of the segments prescribe how to reassemble these exact partial products into the full-precision result.
Concretely, each row of input matrix $A$ is scaled by a power-of-two factor $\mu_i$ to align exponents; Slice $A'_i$ captures bits at positions $\beta i$ through $\beta(i{+}1){-}1$, carrying positional weight $2^{-\beta i}$. The matrix $B$ is split analogously along columns with per-column scales $\boldsymbol{\nu}$:
\begin{equation}
    \begin{aligned}
        A &\;\approx\; \mathrm{diag}(\boldsymbol{\mu})\,
        \sum_{i=0}^{p-1} 2^{-\beta i}\, A'_i\,, \\
        B &\;\approx\; \sum_{j=0}^{p-1} 2^{-\beta j}\, B'_j\, 
        \mathrm{diag}(\boldsymbol{\nu})\,.
    \end{aligned}
\label{eq:scheme-i-split}
\end{equation}
The splitting is \emph{error-free} up to a residual that diminishes with increasing~$p$~\cite{rump-2008, ozaki-1, uchino-2025a}.
These positional weights determine how the slice products are paired. 
The product $A'_i B'_j$ has combined weights $2^{-\beta (i{+}j)}$.
Letting $s = i+j$, all pairs sharing the same $s$ can be accumulated into a single INT32 accumulator $C_s$, yielding $p$~accumulators in a triangular pattern:
\begin{equation}
  C_s \;=\; \sum_{i=0}^{s}
  A'_i \; B'_{s-i}\,,
  \qquad s = 0, \ldots, p{-}1\,,
  \label{eq:scheme-i-compute}
\end{equation}
where each term is a single INT8 GEMM.
Accumulator $C_0$ receives 1~product, $C_1$ receives 2~products, etc., up to~$C_{p-1}$ with~$p$, totaling $p(p{+}1)/2$ INT8 GEMMs.
Each additional slice adds $\sim\!\beta \leq 8$ bits of precision.
The $p$~INT32 accumulators are then merged by a weighted sum we call \emph{shift-reduce}:
\begin{equation}
  C \;=\; \mathrm{diag}(\boldsymbol{\mu})\,
  \Bigl(\sum_{s=0}^{p-1} 2^{-\beta\, s}\, C_s\Bigr)\,
  \mathrm{diag}(\boldsymbol{\nu}).
  \label{eq:scheme-i-reconstruct}
\end{equation}

\subsubsection{\textbf{Scheme~II}}
\label{sec:back-ozaki-ii}

Rather than splitting mantissa bits, Scheme~II maps the computation into modular integer arithmetic~\cite{ozaki-2}.
For any integer~$m$,
\begin{equation}
  (AB) \bmod m \;=\; \bigl((A \bmod m)\,(B \bmod m)\bigr) \bmod m\,,
  \label{eq:mod-mult}
\end{equation}
so the matrix product modulo~$m$ can be computed from the operand residues.
Choosing $m \leq 256$ makes every modular product a single INT8 GEMM.
A single $m$, however, yields only one residue of $AB$; The full integer product is recovered from $p$ residues with pairwise co-prime moduli $m_1, \ldots, m_p$ via the CRT, provided it does not exceed $P = \prod m_\ell$.

To apply this strategy, the input operands are first scaled to integers $A' = \mathrm{trunc}(\mathrm{diag}(\boldsymbol{\mu}) \cdot A)$ and $B' = \mathrm{trunc}(B \cdot \mathrm{diag}(\boldsymbol{\nu}))$ via power-of-two vectors $\boldsymbol{\mu}, \boldsymbol{\nu}$, then reduced modulo $p$ pairwise co-prime moduli $m_1, \ldots, m_p$ (e.g., $m_\ell \in \{256, 255, 253, \ldots\}$):
\begin{equation}
  \label{eq:scheme-ii-mod}
  \begin{aligned}
    A'_\ell &= A' \bmod m_\ell\,, \\
    B'_\ell &= B' \bmod m_\ell\,,
  \end{aligned}
  \qquad \ell = 1,\ldots,p\,.
\end{equation}
Unlike Scheme~I's triangular pairing, each modulus produces exactly one independent INT8 GEMM, which multiplies the residue operands into an INT32 accumulator:
\begin{equation}
  \tilde{C}_\ell \;=\; A'_\ell \, B'_\ell\,.
  \label{eq:scheme-ii-compute}
\end{equation}
A subsequent \emph{modular reduction}, i.e., the element-wise modulo operation $(\cdot) \bmod m_\ell$, collapses each entry of $\tilde{C}_\ell$ into its INT8 residue:
\begin{equation}
  C'_\ell \;=\; \tilde{C}_\ell \bmod m_\ell
  \;\equiv\; A'B' \pmod{m_\ell}\,,
  \label{eq:scheme-ii-reduce}
\end{equation}
which lies in $[0, m_\ell{-}1]$ since $m_\ell \leq 256$.
The full Scheme~II compute phase consists of $p$~INT8 GEMMs followed by $p$~modular reductions, linear in the number of moduli.
The complete integer product~$C' = A'B'$ is then recovered by CRT.
Let $P = \prod_{\ell=1}^{p} m_\ell$ and $q_\ell$ be the modular inverse of $P/m_\ell$ modulo~$m_\ell$. Then
\begin{equation}
  C' \;=\; \sum_{\ell=1}^{p}
  \frac{P}{m_\ell}\, q_\ell \cdot C'_\ell
  \;\bmod\; P.
  \label{eq:crt-reconstruct}
\end{equation}
Reconstruction is exact, provided that $2 \sum_{h} |a'_{ih}|\,|b'_{hj}| < P$ for all~$(i,j)$.
Adding moduli increases $P$, improving the achievable precision to $\log_2 P = \sum_\ell \log_2 m_\ell$~bits.
Finally, the floating-point result is recovered by inverse scaling: $C = \mathrm{diag}(\boldsymbol{\mu})^{-1}\, C'\, \mathrm{diag}(\boldsymbol{\nu})^{-1}$.

\begin{table}[!t]
  \centering
  \caption{Comparison of Ozaki Scheme~I and Scheme~II.}
  \label{tab:scheme-comparison}
  \small
  \begin{tabular}{lcc}
    \toprule
    & \textbf{Scheme~I} & \textbf{Scheme~II} \\
    \midrule
    Decomposition & FP splitting & modular reduction \\
    INT8 GEMMs    & $p(p{+}1)/2$ & $p$ \\
    Scaling       & quadratic    & linear \\
    Reconstruct   & shift-reduce & CRT \\
    Precision gain & $\sim\!8p$ bits & $\log_2 P$ bits \\
    \bottomrule
  \end{tabular}
  \vspace{-1.2em}
\end{table}

Table~\ref{tab:scheme-comparison} summarizes both schemes.
Scheme~I provides coarse-grained precision control (${\sim}\beta$~bits per slice) with quadratic GEMM count.
Scheme~II offers finer granularity ($\log_2 m_\ell$~bits per modulus) at linear cost, but requires a more involved CRT reconstruction.
Despite their algebraic differences, ``naive'' implementations of both schemes are bottlenecked by INT32 intermediate traffic rather than by TC throughput.
Sections~\ref{sec:emu-i} and~\ref{sec:emu-ii} quantify this 
overhead and present fused kernel designs that eliminate it.

\begin{figure*}[t]
    \centering
    \includegraphics[width=\textwidth]{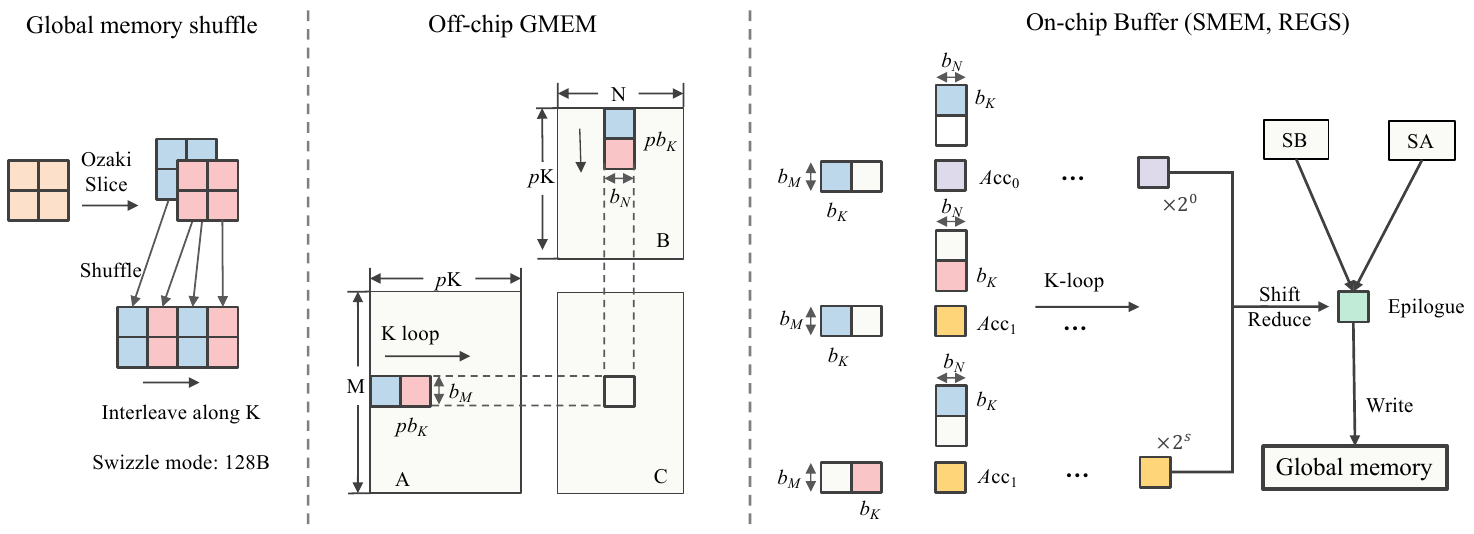}
    \caption{\textbf{End-to-end EmuGEMM-I pipeline.}
    \textbf{Left:} Ozaki decomposition. The input matrices $A$ and $B$ are split into $p$ INT8 slices each. The slices are interleaved along the contraction dimension~$K$ at MMA-tile granularity so that all $p$ slices are fetched together with TMA.
    \textbf{Center:} Off-chip GMEM layout after interleaving. The K-loop streams $K$-chunks of both $A$ and $B$ into shared memory.
    \textbf{Right:} On-chip execution. Each $K$-chunk feeds $p(p{+}1)/2$ MMA into $p$ register-resident accumulators; The epilogue applies shift-reduce
    and writes the reconstructed output tile back to GMEM.}
    \label{fig:pipeline}
    \vspace{-1.2em}
\end{figure*}

\section{EmuGEMM-I: Fused Scheme~I Kernel Design}
\label{sec:emu-i}

In a ``naive'' implementation of Scheme~I, each of the $p(p{+}1)/2$ slice-pair products (Eq.~\eqref{eq:scheme-i-compute}) is issued as a separate GEMM kernel,
and a final reconstruction kernel reads the $p$ INT32 partial sums from GMEM to assemble the result in the target precision. 
The end-to-end GMEM traffic is
\begin{equation}
  T_{\text{naive}}
    = \underbrace{\tfrac{p(p{+}1)}{2}\,(M{+}N)\,K}_%
      {\text{operand loads}}
    + \underbrace{4p(p{+}1)\,\,M N }_%
      {\text{INT32 traffic}}
    + \underbrace{b\,M N}_%
      {\text{final output}}\,,
  \label{eq:traffic-naive}
\end{equation}
where $b$ is the size of the output element (4 for FP32, 8 for FP64).
The INT32 term aggregates $4p^2 MN$ accumulator writes during the compute phase and $4p\,MN$ partial-sum reads during reconstruction.
If, instead, all slice products are \emph{fused} into one kernel where each operand slice is loaded from GMEM exactly once and all $p$ accumulators reside on-chip, the traffic reduces to
\begin{equation}
  T_{\text{fused}}
    = \underbrace{p\,(M{+}N)\,K}_%
      {\text{operand loads (once per slice)}}
    + \underbrace{b\,M N}_%
      {\text{final output}}\,.
  \label{eq:traffic-fused}
\end{equation}
The entire second term of Eq.~\eqref{eq:traffic-naive} vanishes: $p$~INT32 accumulators never leave on-chip storage.
Since each INT32 element is $4\times$ larger than an INT8 element, keeping the accumulators on the chip is even more valuable than reducing operand loads. The resulting arithmetic intensity increases by a factor of $(p{+}1)/2$. For FP64 emulation (i.e., $p{=}8$), a $4.5\times$ increase pushes the kernel deeper into the compute-bound region.

This fusion, however, comes at a cost: It multiplies both the on-chip accumulator~$\mathrm{Acc}$ and the SMEM operand buffer~$S_{\mathrm{op}}$ (Table~\ref{tab:gemm-model}) by~$p$.
With fixed per-SM budgets, smaller tiles or shallower pipelines are needed, both of which degrade throughput (Section~\ref{sec:back-gemm}).
Moreover, with $p$~separate matrices per operand, each $K$-step would expect $p$~independent TMA loads per operand from non-contiguous addresses, increasing scheduling complexity and SMEM management overhead.
Achieving fusion, therefore, requires co-designing two components:
(1)~a data layout that allows for all $p$~slices to be streamed efficiently via TMA without extra SMEM, and
(2)~a kernel structure that executes $p(p{+}1)/2$ MMA instructions per $K$-step while fitting $p$~accumulators on-chip.
We address them in the following subsections.

\subsection{Interleaved Data Layout}
\label{sec:emu-i-layout}

The data layout must enable delivery of all $p$~slices to SMEM via a single TMA descriptor per $K$-step, while simultaneously satisfying three on-chip constraints.
First, all $p$~INT32 accumulators must remain on-chip throughout the $K$-loop, consuming $p \cdot \mathrm{Acc}$ bytes (RF on Hopper, TMEM on Blackwell).
Second, the $p$~operand slices must fit in SMEM: Each $K$-step buffers $p\,(b_M{+}b_N)\,b_K\,\gamma$ bytes for $\gamma$-stage pipelining, where the factor~$\gamma$ accounts for multi-stage prefetching to hide TMA latency.
Third, each slice must occupy an MMA-aligned offset so that the triangular schedule (Eq.~\eqref{eq:scheme-i-compute}) can index operands via compile-time constants.
We satisfy all three constraints by \emph{interleaving} the $p$~slices of each operand along the $K$-dimension at $t_K$ granularity during preprocessing.

Figure~\ref{fig:pipeline} illustrates the resulting layout and execution pipeline.
Concretely, the decomposition kernel writes each slice's $t_K$-wide column block directly to its interleaved position in GMEM in a single preprocessing pass: for slice $i \in [0,p)$ and $K$-chunk $c \in [0, K/t_K)$,
\begin{align}\label{eq:interleave}
  \hat{A}[\,:\,,\; &(cp{+}i)\,t_K : (cp{+}i{+}1)\,t_K\,]
  \nonumber\\
  &= A'_i[\,:\,,\; ct_K : (c{+}1)t_K\,]\,.
\end{align}
For example, with $p{=}3$ slices and $K$-chunks $c = 0, 1, \ldots$, the columns of $\hat{A}$ are laid out as $[A'_0\,|\,A'_1\,|\,A'_2\,|\,A'_0\,|\,A'_1\,|\,A'_2\,|\,\cdots]$, each block $t_K$~columns wide.
Because the interleaving granularity matches the MMA $K$-tile width~$t_K$, each slice begins at a tile-aligned boundary in SMEM, satisfying the third constraint without additional padding or remapping.
The result is a single matrix $\hat{A}$ of shape $M \times pK$: Consecutive $t_K$-wide column groups cycle through slices $0, 1, \ldots, p{-}1$ before advancing to the next $K$-chunk (Fig.~\ref{fig:pipeline}, left). Operand~$B$ is interleaved analogously along its row dimension, resulting in $\hat{B}$ of shape $pK \times N$.

From TMA's perspective, $\hat{A}$ and $\hat{B}$ are ordinary 2D matrices. Each $K$-loop iteration loads a contiguous tile of shape $b_M \times p\,b_K$ (for~$A$) or $p\,b_K \times b_N$ (for~$B$) via a single TMA descriptor per operand (Fig.~\ref{fig:pipeline}, center). Within SMEM, slice~$s$ sits at a fixed offset from the tile base, directly addressable by the MMA instruction without any runtime data movement. 
Together with the MMA-aligned offsets within each tile, a single layout transformation addresses all co-design requirements.

At runtime, our kernel operates identically to a standard tiled GEMM, with the sole difference that the effective $K$-dimension is $pK$ instead of~$K$, proportionally increasing the number of $K$-loop iterations and the operand footprint in GMEM.
This layout is agnostic to the slice datatype: It applies equally to INT8 (this work), FP16/BF16 for FP32 emulation, and any future format with a fixed MMA $K$-tile width~$t_K$.

\subsection{Fused Persistent Kernel}

\begin{algorithm}[t]
\DontPrintSemicolon
\caption{Fused Scheme~I GEMM kernel.}\label{alg:fused-gemm}
\KwIn{$\hat{A} \in \mathbb{Z}_{\mathrm{INT8}}^{M \times pK}$,
      $\hat{B} \in \mathbb{Z}_{\mathrm{INT8}}^{pK \times N}$
      (interleaved)}
\KwOut{$C \in \mathbb{R}^{M \times N}$ in GMEM}
\For{each $b_M \times b_N$ output tile $(m, n)$}{
  $C_{\mathrm{acc}}[0 \ldots p{-}1] \leftarrow 0$\;
  \For{$k \leftarrow 0$ \KwTo
       $K{-}b_K$ \textbf{by} $b_K$}{
    $A_{\mathrm{smem}} \leftarrow
        \textsc{TMA}(\hat{A}[\ldots,\;
        p\,k\!:\!p(k{+}b_K)])$\;
    $B_{\mathrm{smem}} \leftarrow
        \textsc{TMA}(\hat{B}[
        p\,k\!:\!p(k{+}b_K),\;\ldots])$\;
    \For{$s \leftarrow 0$ \KwTo $p{-}1$}{
      \For{$i \leftarrow 0$ \KwTo $s$}{
        $C_{\mathrm{acc}}[s] \mathrel{+}=
            \textsc{MMA}(
            A_{\mathrm{smem}}[i],\,
            B_{\mathrm{smem}}[s{-}i])$\;
      }
    }
  }
  $C_{\mathrm{fp}} \leftarrow 0$\;
  \For{$s \leftarrow 0$ \KwTo $p{-}1$}{
    $C_{\mathrm{fp}} \mathrel{+}=
        2^{-\beta s} \cdot
        \mathrm{FP}(C_{\mathrm{acc}}[s])$\;
  }
  $C[m\!:\!m{+}b_M,\; n\!:\!n{+}b_N]
      \leftarrow
      \mathrm{diag}(\boldsymbol{\mu})\,
      C_{\mathrm{fp}}\,
      \mathrm{diag}(\boldsymbol{\nu})$\;
}
\end{algorithm}

With the interleaved layout delivering all $p$~slices into SMEM via a single TMA load, the kernel must now execute the $p(p{+}1)/2$ slice-pair products and the shift-reduce epilogue entirely on-chip (Fig.~\ref{fig:pipeline}, right).
Algorithm~\ref{alg:fused-gemm} summarizes the resulting design.

\textbf{Triangular scheduling.}
Each $K$-step begins with a single TMA load that brings $p$~slices of both operands $\hat{A}$  (Eq.~\eqref{eq:interleave}) and $\hat{B}$ into SMEM (lines~4--5).
The key observation is that once all $p$~slices reside in SMEM, the $p(p{+}1)/2$ products $A'_i B'_j$ (Eq.~\eqref{eq:scheme-i-compute}) can be computed by issuing MMA instructions directly on the buffered data without any additional data movement.
Recall from Section~\ref{subsubsec:back-ozaki-i} that the accumulator~$C_s$ collects the $s{+}1$ products $A'_i B'_{s-i}$ with $i = 0, \ldots, s$, all sharing positional weight $2^{-\beta s}$ (Eq.~\eqref{eq:scheme-i-compute}).
The inner loop (lines~6--8) iterates accordingly: For each $s = 0, \ldots, p{-}1$, it issues MMA instructions for $A'_i \cdot B'_{s-i}$ with $i = 0, \ldots, s$, accumulating each result into $C_{\mathrm{acc}}[s]$. 
Furthermore, the above layout places each slice at a fixed offset within the SMEM tile, allowing the compiler to resolve operand addresses at compile time  (Fig.~\ref{fig:pipeline}, right).
The remaining kernel infrastructure is inherited from the standard tiled GEMM design (Section~\ref{sec:back-gemm}).

\textbf{Shift reduce epilogue.}
After the $K$-loop completes, the $p$-INT32 accumulators reside entirely in RF (Hopper) or TMEM (Blackwell). 
The epilogue applies the shift-reduce reconstruction of Eq.~\eqref{eq:scheme-i-reconstruct} (lines~10--12): Each accumulator $C_{\mathrm{acc}}[s]$ is converted into floating-point format, weighted by $2^{-\beta s}$, where $\beta \leq 8$ is the per-slice bit-width from the Ozaki decomposition (Section~\ref{subsubsec:back-ozaki-i}), and the $p$~weighted tiles are summed into a single FP32 or FP64 output.
This reconstruction is performed entirely on-chip without intermediate SMEM or GMEM traffic.
Because the base-$2^\beta$ weights are exact powers of two, the shift-reduce introduces no rounding error beyond the decomposition residual established in Section~\ref{subsubsec:back-ozaki-i}.
The epilogue's SMEM footprint remains $S_{\mathrm{epi}} = 4\,b_M b_N$ bytes, identical to a standard GEMM.
Because the kernel is persistent, TMA already prefetches the next output tile's operands while the current tile's shift-reduce drains through the RF, overlapping epilogue computation with data movement.

\textbf{Performance trade-off.}
The fusion described above eliminates all intermediate traffic but increases the on-chip storage requirements.
In particular, fusing $p$~slices scales the SMEM operand buffer to $S_{\mathrm{op}}^{(p)} = p \cdot S_{\mathrm{op}}$ and the on-chip accumulator to $\mathrm{Acc}^{(p)} = 4\,p\,\alpha\,t_M\,t_N$ bytes (the epilogue buffer is independent of~$p$).
Since $\mathrm{Acc}^{(p)}$ must not exceed the per-SM accumulator capacity $\mathrm{Acc}_{\max}$, the maximum M-stacking factor is
\begin{equation}\label{eq:alpha-max}
  \alpha_{\max}
    = \left\lfloor
        \frac{\mathrm{Acc}_{\max}}{4\,p\,t_M\,t_N}
      \right\rfloor\,.
\end{equation}
Given that each $K$-step issues $\omega = \alpha \cdot \sigma$ MMA instructions (Section~\ref{sec:back-gemm}), the triangular schedule multiplies this by $p(p{+}1)/2$, yielding an effective pipeline depth of
\begin{equation}\label{eq:omega-eff}
  \omega_{\mathrm{eff}} = \frac{p(p{+}1)\omega}{2}\,.
\end{equation}
Even though increasing~$p$ reduces~$\alpha_{\max}$ and thus~$\omega$, the quadratic growth of the triangular factor ensures that $\omega_{\mathrm{eff}}$ remains well above the saturation threshold of~16 (Section~\ref{sec:back-gemm}).

On Hopper, the accumulator and thread-local state (epilogue arithmetic, address computation) share the same RF, so increasing~$p$ directly reduces the available RF budget, limiting $p \times \alpha$.
On Blackwell, MMA output is written directly to the dedicated TMEM, while RF is reserved for thread-local state only.
During the epilogue, accumulator tiles are drained from TMEM into RF in a pipelined fashion for shift-reduce computation, so only a subset of the $p$~accumulators resides in RF.
This separation yields a substantially larger effective on-chip
budget than on Hopper at any given slice count.

\section{EmuGEMM-II: Fused Scheme~II Kernel Design}
\label{sec:emu-ii}

\begin{figure}[t]
    \centering
    \vspace{-0.6em}
    \includegraphics[width=\columnwidth]{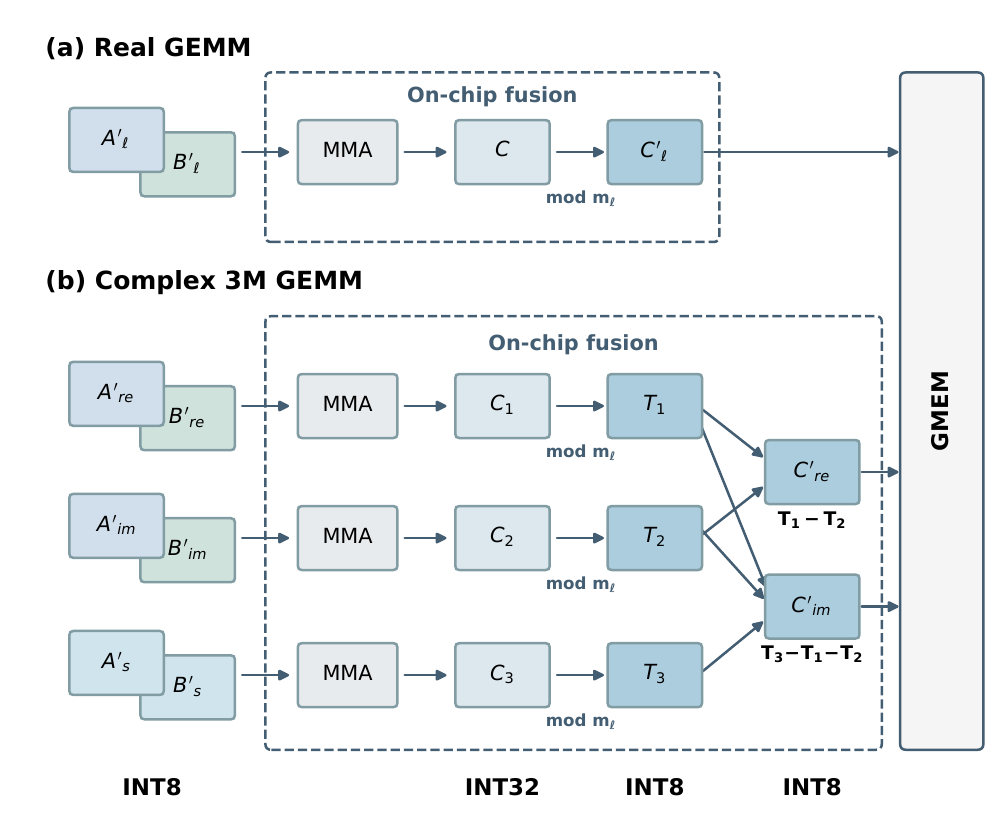}
    \caption{\textbf{EmuGEMM-II kernel structure} for one 
    modulus~$m_\ell$.
    \textbf{(a)}~Real GEMM: A single INT8 MMA produces an 
    INT32 accumulator~$C$, which is reduced to INT8 via 
    $\bmod\,m_\ell$ on-chip before writing to GMEM, 
    eliminating the $4{\times}$ write amplification of a 
    naive INT32 round-trip.
    \textbf{(b)}~Complex 3M GEMM: Three independent MMA 
    instructions produce INT32 accumulators $C_1$, $C_2$, 
    $C_3$, each reduced to INT8 ($T_1$, $T_2$, $T_3$) 
    on-chip.
    The 3M combination ($C'_{re} = T_1 - T_2$, 
    $C'_{im} = T_3 - T_1 - T_2$) is then computed in INT8 
    and written to GMEM.
    }
    \label{fig:emu-ii-pipeline}
    \vspace{-1.4em}
\end{figure}

Unlike Scheme~I, where fusion targets operand reuse across $p(p{+}1)/2$ triangular products (Eq.~\eqref{eq:scheme-i-compute}), Scheme~II issues only $p$~independent GEMMs (Eq.~\eqref{eq:scheme-ii-compute}), one per modulus.
Each operand slice is already loaded exactly once, so operand-side traffic is inherently linear in~$p$.
The fusion opportunity, instead, lies on the output side: 
A ``naive'' implementation writes each INT32 accumulator $\tilde{C}_\ell$ to GMEM only to read it back for the modular reduction of Eq.~\eqref{eq:scheme-ii-reduce}.
Focusing on the computed phase traffic per modulus, the dominant terms are
\begin{equation}
  T_{\text{naive}}
    = \underbrace{(M{+}N)\,K}_%
      {\text{operand loads}}
    + \underbrace{8\,M N}_%
      {\text{INT32 round-trip}}
    + \underbrace{M N}_%
      {\text{INT8 write}}\,.
  \label{eq:traffic-ii-naive}
\end{equation}
If instead the modular reduction is performed in-register within the GEMM epilogue, the kernel writes only the INT8 residue:
\begin{equation}
  T_{\text{fused}}
    = \underbrace{(M{+}N)\,K}_%
      {\text{operand loads}}
    + \underbrace{M N}_%
      {\text{INT8 write}}\,.
  \label{eq:traffic-ii-fused}
\end{equation}
The reduction in output traffic, $8{\times}$, directly translates into increased arithmetic intensity.

\subsection{In-Register Modular Reduction}
\label{subsec:emu-ii-mod}

For real GEMM, the fusion mechanism is straightforward (Fig.~\ref{fig:emu-ii-pipeline}(a)).
After the $K$-loop accumulates the full INT32 product $C_{\mathrm{int32}} = A'_\ell B'_\ell$, the epilogue computes $C'_\ell = C_{\mathrm{int32}} \bmod m_\ell$ element-wise using the GPU's native integer modulo instruction, converts the result to INT8, and writes it to GMEM.
This modular reduction is lightweight: The moduli are compile-time integer constants, and the resulting residues fit in INT8, so each element requires only a few integer instructions.
As the kernel is persistent, TMA prefetches the next tile's operands while the current tile's epilogue executes, overlapping the reduction cost with data 
movement.
Hence, the kernel produces the same INT8 residue matrix $C'_\ell$ as the naive implementation, but without the INT32 round-trip to GMEM.
The kernel structure is, otherwise, identical to a standard INT8 GEMM (Section~\ref{sec:back-gemm}), a direct consequence of Scheme~II's linear GEMM count: With only one product per modulus, there is no need for the interleaved layout or triangular scheduling as in Scheme~I.

\subsection{Fused 3M Complex Kernel}

For complex operands $A = A_{re} + i\,A_{im}$ and $B = B_{re} + i\,B_{im}$, the modular decomposition of Section~\ref{sec:back-ozaki-ii} is applied independently to the real and imaginary parts, producing four INT8 
residue matrices $A'_{\ell,re}$, $A'_{\ell,im}$, $B'_{\ell,re}$, $B'_{\ell,im}$ per modulus.
The standard 4M formulation computes $C'_{\ell,re}$ and $C'_{\ell,im}$ from four real products $A'_{\ell,re}B'_{\ell,re}$, $A'_{\ell,im}B'_{\ell,im}$, $A'_{\ell,re}B'_{\ell,im}$, and $A'_{\ell,im}B'_{\ell,re}$.
The 3M identity~\cite{uchino-2025c} reduces this to three products at the cost of one extra addition on each operand, using the sum $A'_{\ell,\Sigma} \equiv A'_{\ell,re} + A'_{\ell,im}$ (and analogously for $B$):
\begin{equation}
  \begin{aligned}
    T_1 &= A'_{\ell,re}\,B'_{\ell,re}\,,\qquad
    T_2 = A'_{\ell,im}\,B'_{\ell,im}\,, \\
    T_3 &= A'_{\ell,\Sigma}\,B'_{\ell,\Sigma}\,.
  \end{aligned}
  \label{eq:scheme-ii-complex-3m}
\end{equation}
The complex output is reconstructed as $C'_{\ell,re} = T_1 - T_2$ and $C'_{\ell,im} = T_3 - T_1 - T_2$.
In native floating-point arithmetic, the 3M identity suffers catastrophic cancellation in $T_3 - T_1 - T_2$ when the real and imaginary parts have similar magnitude~\cite{higham-1992}.
In modular integer arithmetic, all operations are exact: No rounding occurs, and the subtraction is computed without such error.
This makes 3M strictly preferable for Scheme~II, reducing GEMM counts by 25\% without any accuracy penalty.

\textbf{Naive baseline.}
In a ``naive'' implementation, each of the three products $T_1$, $T_2$, $T_3$ is launched as a separate GEMM kernel, each writing an INT32 accumulator to GMEM.
A follow-up kernel then reads these three INT32 matrices back and reduces them to INT8 via $\bmod\,m_\ell$, as in the real-valued case (Section~\ref{subsec:emu-ii-mod}).
The 3M combination ($C'_{\ell,re} = T_1 - T_2$, $C'_{\ell,im} = T_3 - T_1 - T_2$) is computed in INT8 and written to GMEM.
The per-modulus traffic is
\begin{equation}
  T_{\text{naive,3M}}
    = \underbrace{3\,(M{+}N)\,K}_%
      {\text{operand loads}}
    + \underbrace{24\,M N}_%
      {\text{INT32 round-trips}}
    + \underbrace{2\,M N}_%
      {\text{INT8 writes}}\,.
  \label{eq:traffic-ii-naive-3m}
\end{equation}
The $24\,MN$ intermediate term (three INT32 matrices, each written and read back) dominates the output traffic.

\textbf{Fused kernel.}
The fused 3M kernel eliminates all intermediate materialization by computing all three products within a single persistent kernel per modulus (Fig.~\ref{fig:emu-ii-pipeline}(b)).
The kernel first executes a complete $K$-loop for $T_1$ ($A'_{re} \times B'_{re}$), accumulating into a single INT32 accumulator.
After this $K$-loop, the epilogue reduces the accumulator to INT8 via $\bmod\,m_\ell$ and stores the result on-chip.
The same accumulator is then reused for $T_2$ ($A'_{im} \times B'_{im}$), whose INT8 result is similarly stored on-chip after reduction.
Finally, $T_3$ ($A'_{\Sigma} \times B'_{\Sigma}$) is computed in a third $K$-loop pass.
At this point all three INT8 values $T_1$, $T_2$, $T_3$ are available on-chip, and the 3M combination $C'_{\ell,re} = T_1 - T_2$, $C'_{\ell,im} = T_3 - T_1 - T_2$ is computed in INT8 before writing only two INT8 tiles to GMEM.
This sequential design keeps only one INT32 accumulator on-chip at any time, so the accumulator footprint is identical to a single real GEMM.
The two previously reduced INT8 results require negligible additional storage compared to the INT32 accumulator they replace.
The aggregate compute-phase traffic for the fused 3M kernel 
per modulus is
\begin{equation}
  T_{\text{fused,3M}}
    = \underbrace{3\,(M{+}N)\,K}_%
      {\text{operand loads}}
    + \underbrace{2\,M N}_%
      {\text{INT8 writes}}\,.
  \label{eq:traffic-ii-fused-3M}
\end{equation}
The operand loads are identical to the naive case; The entire $24\,MN$ intermediate term vanishes.
Compared with three independent real fused GEMMs (Eq.~\eqref{eq:traffic-ii-fused}), which would write $3\,MN$ bytes of INT8 output, the 3M kernel writes only $2\,MN$ since the intermediate results, $T_1$ and $T_2$, are retained on the chip and never materialized to GMEM.

\subsection{Reconstruction and Resource Analysis}
The CRT reconstruction (Eq.~\eqref{eq:crt-reconstruct}) is implemented via a separate kernel, as the multi-word integer arithmetic required by the CRT coefficients exceeds the register budget when fused with the GEMM kernel.
Turning to the fused GEMM kernel itself, the sequential design ensures that the accumulator footprint per CTA is $\mathrm{Acc} = 4\,\alpha\,t_M\,t_N$ bytes, identical to a single real GEMM and \emph{independent of~$p$}: Adding moduli does not shrink $\alpha_{\max}$ nor reduce the pipeline depth~$\omega$.
This is the key contrast with EmuGEMM-I, where the $p$-fold accumulator scaling forces progressively smaller tiles at higher precision.

\section{Evaluation}
\label{sec:evaluation}

\subsection{Implementation}
\label{sec:eval-impl}

We develop our optimized implementations of the fused Ozaki Schemes~I and II on top of DeepGEMM~\cite{deepgemm}, 
a lightweight GEMM library that provides abstractions over NVIDIA's compute unified tensor (CuTe) algebra~\cite{cecka2026cutelayoutrepresentationalgebra} and its CUTLASS~\cite{Thakkar_CUTLASS_2023} implementation.
DeepGEMM avoids the heavily templated syntax of the latter and simplifies the development of the fused kernels presented in Sections~\ref{sec:emu-i}--\ref{sec:emu-ii}.
We extend the library's runtime with INT8 operand support and INT32 accumulation, which DeepGEMM does not natively provide.
Appropriate values for the $(\alpha, t_N, \beta, \gamma)$ parameters are selected via DeepGEMM's integrated autotuner.

\subsection{Experimental Setup}
\label{sec:eval-setup}

We run benchmarks on two GPUs, an NVIDIA GH200 (Hopper) GPU, with 96\,GiB HBM3 at up to 4\,TB/s bandwidth, delivering 67\,Tflop/s in FP64 and 1{,}979\,Top/s in INT8,
and an NVIDIA B200 GPU (Blackwell), provisioned through Vast.ai~\cite{vastai}, with 192\,GiB HBM3e at up to 8\,TB/s bandwidth, 40\,Tflop/s in FP64 and 4{,}500\,Top/s in INT8.
All tested kernels, including baselines, run entirely on a single GPU. The host configuration does not affect the reported timings. All experiments use CUDA~13.1 on Ubuntu~24.04 LTS, with GCC~13.4.0 as \texttt{nvcc}'s host compiler.

We compare our kernels against cuBLAS in both native mode (FP64, FP32, TF32, INT8) and Ozaki Scheme~I
emulation~\cite{schwarz2026guaranteed}. For the latter, we disable automatic dynamic precision (ADP) and explicitly select the slice count, enabling direct performance and accuracy comparisons.
We also compare to the GEMMul8 library~\cite{gemmul8}, which implements Ozaki Scheme~II for both real and complex GEMM, including the 3M formulation.
The GEMMul8 algorithms exist in two variants, one focused on numerical accuracy (GEMMul8-Accu) and one prioritizing execution speed (GEMMul8-Fast).
The input matrices are square ($M{=}N{=}K$). The exact sizes are specified per figure.
Each entry is sampled as
\begin{equation}
  a_{ij} = (\mathrm{rand}-0.5)\,
           \exp\,(\phi \cdot \mathrm{randn}),
  \label{eq:test-matrix}
\end{equation}
with conditioning parameter $\phi{=}4.0$, following~\cite{uchino-2025b}.
Larger~$\phi$ widens the exponent range and increases the number of slices needed for a given precision.
Unless specified otherwise, we measure ``effective performance'' as the ratio of the reference matrix-multiplication workload, $2N^3$ for real and $8N^3$ for complex GEMM (regardless of implementation), to the execution runtime.
Except for the kernel-only analysis of Section~\ref{sec:eval-efficiency}, all reported throughputs are end-to-end and include the decomposition and reconstruction stages, i.e., they reflect the performance delivered when EmuGEMM is used as a complete GEMM replacement.
We also measure, for each data point, the relative error with respect to the reference product (standard cuBLAS D/ZGEMM), expressed as ``effective bits of precision,'' i.e., the absolute value of the error's exponent in base-2. For example, a relative error of $10^{-15} \approx 2^{-50}$ corresponds to ${\sim}50$ bits of precision.
All reported measurements are median values of at least 30 executions.
The dispersion around these medians is small: The typical interquartile range relative to the median is about 0.4\% on the GH200 and 0.8\% on the B200, and 95\% of the samples are within 10\% of the median across all data points.

\begin{figure}[t]
    \centering
    \includegraphics[width=\columnwidth]{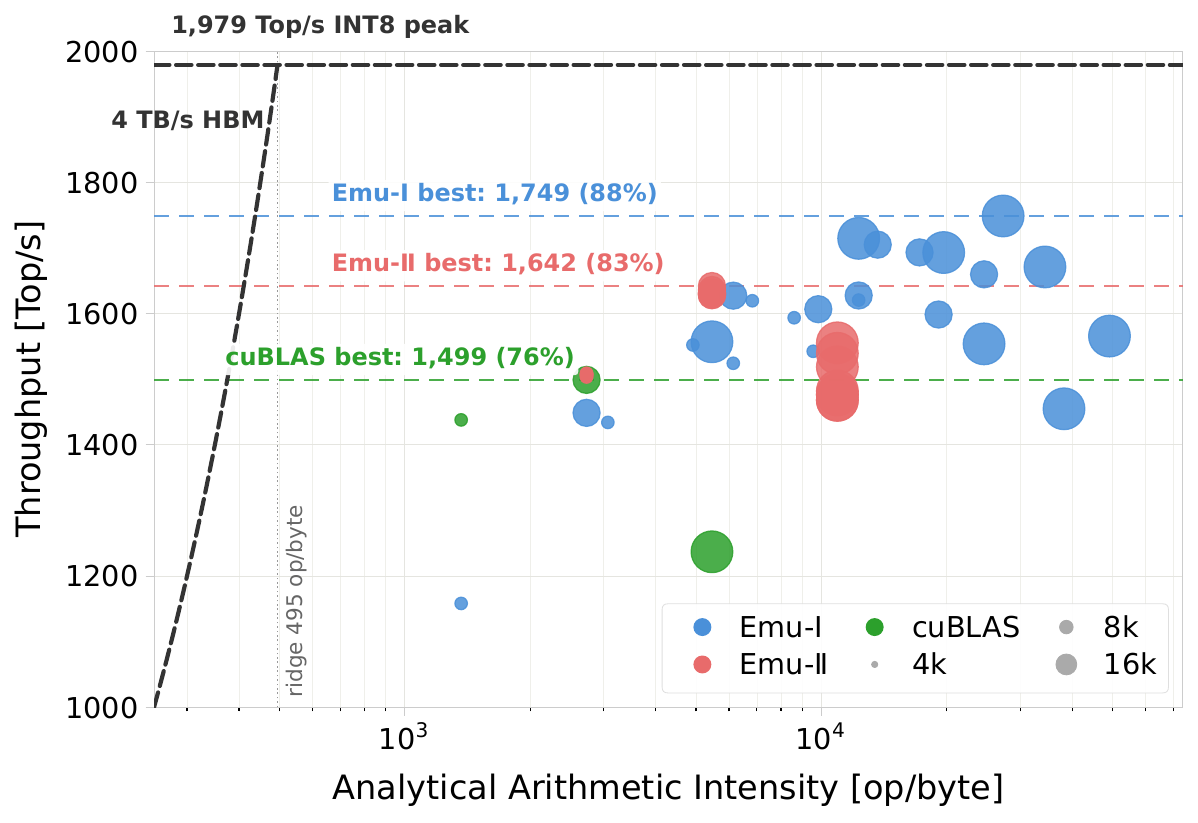}
    \caption{Mapping of EmuGEMM-I and EmuGEMM-II on the GH200 INT8 throughput--intensity plane (analytical arithmetic intensity). The dashed lines mark the 1{,}979~Top/s INT8 and 4~TB/s HBM ceilings. They meet at the 495~op/byte ridge. Measurements cover $p\,{\in}\,\{1,\ldots,8\}$ (Scheme~I) and $p\,{\in}\,\{8,\ldots,15\}$ (Scheme~II) over $M{=}N{=}K \in \{4096,\,8192,\,16384\}$. The bubble area corresponds to the matrix dimension.}
    \label{fig:roofline}
    \vspace{-1.0em}
\end{figure}

\subsection{Fused Kernel Efficiency}
\label{sec:eval-efficiency}

Sections~\ref{sec:emu-i} and~\ref{sec:emu-ii} argued that eliminating the INT32 round-trip through GMEM is the core optimization shared by both EmuGEMM variants. We now validate this argument on hardware. Figure~\ref{fig:roofline} places EmuGEMM-I, EmuGEMM-II, and cuBLAS native INT8 GEMM on a ``throughput versus arithmetic intensity'' plane. 
All throughputs are kernel-only, excluding the slicing and reconstruction stages. The relevant reference is cuBLAS native INT8 GEMM: ``Naive'' Scheme~I or Scheme~II implementations perform their per-slice products with this kernel. Hence, its highest throughput is the best achievable performance for any non-fused emulation kernel.
EmuGEMM-I reaches 1{,}749~Top/s (88\% of INT8 peak), and EmuGEMM-II 1{,}642~Top/s (83\%). Both exceed cuBLAS native INT8 (1{,}499~Top/s, 76\%) by 17\% and 10\%, respectively. Both fused kernels run faster than the ceiling of any naive emulation, highlighting the benefit of the kernel-level fusion argument in Sections~\ref{sec:emu-i} and~\ref{sec:emu-ii}. Removing the INT32 round-trip provides throughput that no naive implementation can reach, regardless of engineering effort spent on the per-slice kernel.
We also report that EmuGEMM-I achieves end-to-end 1{,}639 and 3{,}654 Top/s on GH200 and B200, respectively. Notably, the slicing and reconstruction overheads shrink with increasing matrix size~$N$.

\begin{figure*}[!t]
    \centering
    \includegraphics[width=\textwidth]{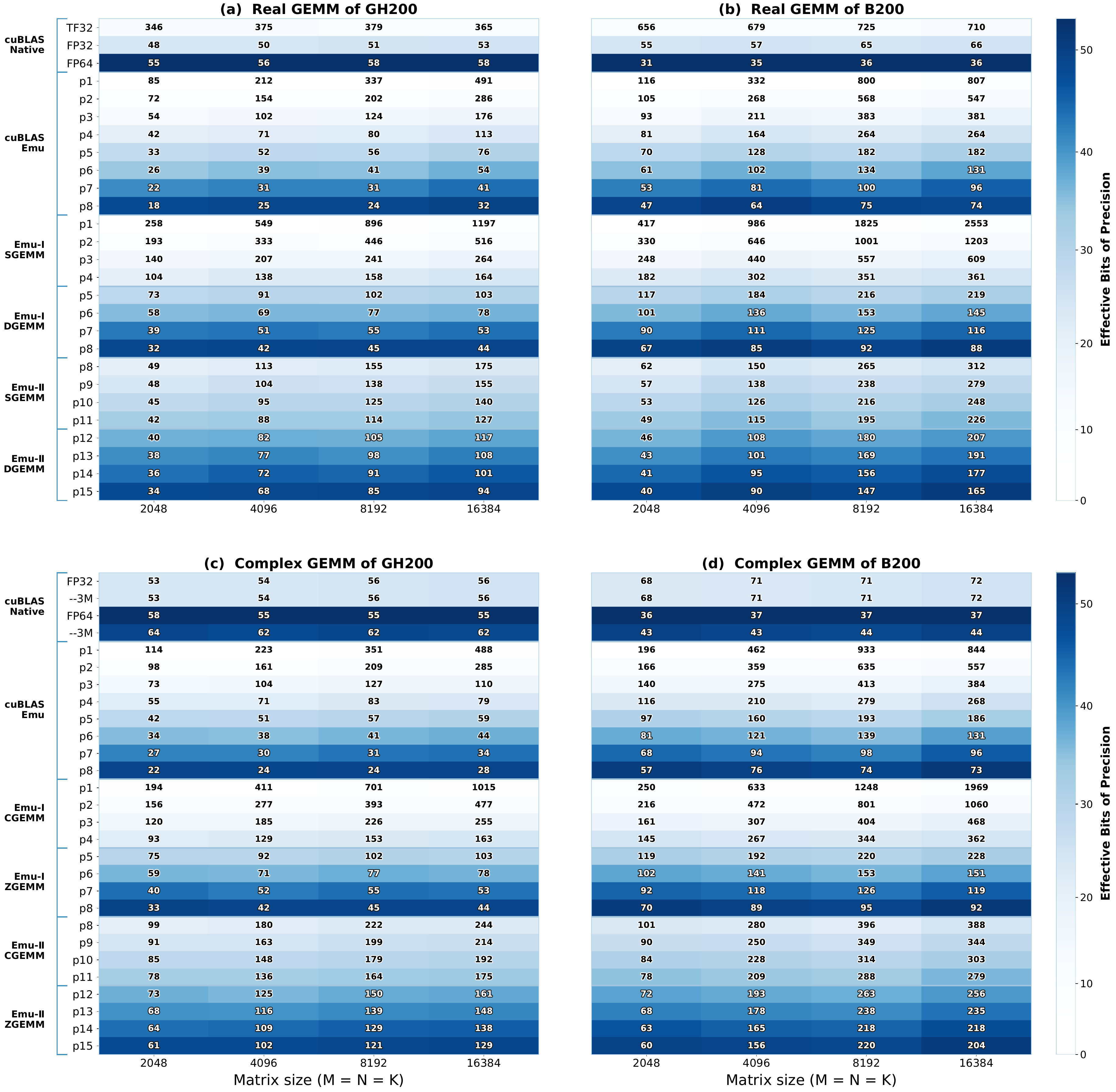}
    \vspace{-0.2em}
    \caption{\textbf{End-to-end effective throughput (cell text, Tflop/s) and precision (cell color, bits, see scale on the right)} of real GEMM (top row: a,\,b) and complex GEMM (bottom row: c,\,d) on NVIDIA GH200 (left column) and B200 (right column). The numeric suffix of the row labels (p1--p15) indicates the number of slices used in Scheme~I and Scheme~II kernels. Panels~(a) and~(b) compare cuBLAS native baselines (TF32/FP32/FP64), cuBLAS Scheme-I emulation (p1--p8), EmuGEMM-I SGEMM (p1--p4), EmuGEMM-I DGEMM (p5--p8), EmuGEMM-II SGEMM (p8--p11), and EmuGEMM-II DGEMM (p12--p15). Panels~(c) and~(d) extend to complex arithmetic with cuBLAS native CGEMM/ZGEMM (with and without the 3M identity), EmuGEMM-I CGEMM/ZGEMM, and EmuGEMM-II CGEMM/ZGEMM.}
    \label{fig:heatmap}
    \vspace{-1.6em}
\end{figure*}

\subsection{End-to-End Performance Across Configurations}
\label{sec:eval-results}

Section~\ref{sec:eval-efficiency} isolated the fused kernel itself. We now widen the lens to the full precision--performance spectrum of EmuGEMM.
Figure~\ref{fig:heatmap} reports end-to-end effective Tflop/s over Schemes~I and II, GH200 and B200 GPUs, real and complex GEMM, and a range of matrix sizes and slice counts~$p$. Cell color encodes the effective bits of precision delivered, and the caption describes the layout and baselines.
Rows in each panel are ordered by increasing precision, so EmuGEMM sweeps the full range from TF32-level down to full FP64, in contrast with the three native operating points (TF32, FP32, FP64).
We also compare against cuBLAS Scheme~I emulation~\cite{schwarz2026guaranteed}. For complex GEMM, EmuGEMM-I uses the 4M formulation (as does cuBLAS emulation), while EmuGEMM-II takes advantage of 3M, matching Ozaki's canonical Scheme~II complex variant. 

\textbf{Low-precision region.}
At every point up to single-precision target accuracy, EmuGEMM-I exceeds both cuBLAS native and Scheme~I emulation. At $p{=}4$ and $N{=}4{,}096$ on the GH200, EmuGEMM-I delivers 138~Tflop/s, a $1.9\times$ speedup over cuBLAS Scheme~I emulation at the same~$p$ (71~Tflop/s) and a $2.8\times$ speedup over cuBLAS native FP32 (50~Tflop/s). At $N{=}16{,}384$, two-slice EmuGEMM-I reaches 516~Tflop/s on the GH200 and 1{,}203~Tflop/s on the B200, $1.4\times$ and $1.7\times$ over cuBLAS native TF32 (365 and 710~Tflop/s), while delivering similar precision to TF32 in both cases. Complex GEMM shows the same ranking: at $p{=}4, N{=}4{,}096$ on the GH200, EmuGEMM-I CGEMM reaches 129~Tflop/s, $2.4\times$ over cuBLAS native FP32 CGEMM (54~Tflop/s).

\textbf{High-precision region.}
As the target precision approaches FP64, Scheme~II becomes the better choice. Scheme~I at its highest slice counts ($p\,{\in}\,\{7,8\}$) cannot match cuBLAS native FP64 on the GH200, where Hopper still provides abundant dedicated FP64 silicon: At $p{=}8, N{=}16{,}384$, EmuGEMM-I reaches 44~Tflop/s, cuBLAS Scheme~I emulation 32~Tflop/s, while cuBLAS native FP64 achieves 58~Tflop/s. Neither Scheme~I implementations can close this gap within the algorithm's $p(p{+}1)/2$ compute budget. This is precisely the motivation behind the development of Scheme~II. At the maximum tested precision ($p{=}15$) and $N{=}16{,}384$, EmuGEMM-II delivers 94~Tflop/s on the GH200 and 165~Tflop/s on the B200, $1.6\times$ and $4.6\times$ more than cuBLAS native FP64 (58 and 36~Tflop/s). On complex GEMM, the advantage is even larger: EmuGEMM-II ZGEMM at the same $(p, N)$ reaches 129~Tflop/s on the GH200 and 204~Tflop/s on the B200, $2.3\times$ and $5.5\times$ over cuBLAS native ZGEMM (55 and 37~Tflop/s).

\textbf{Emulation gains on Blackwell.}
The cuBLAS native FP64 throughput drops from 58 to 36~Tflop/s between GH200 and B200, mirroring the reduced FP64 silicon share of the latter hardware. EmuGEMM-II moves in the opposite direction: $p{=}15$ throughput rises from 94 to 165~Tflop/s on the same matrices. Architectures that concentrate area on low-precision TCs favor emulation strongly, and the trend is likely to continue on future AI-optimized GPU generations. Complex GEMM follows the same cross-architecture pattern.

\begin{figure*}[t]                                            
    \centering   
    \vspace{-1em}
    \includegraphics[width=\textwidth]{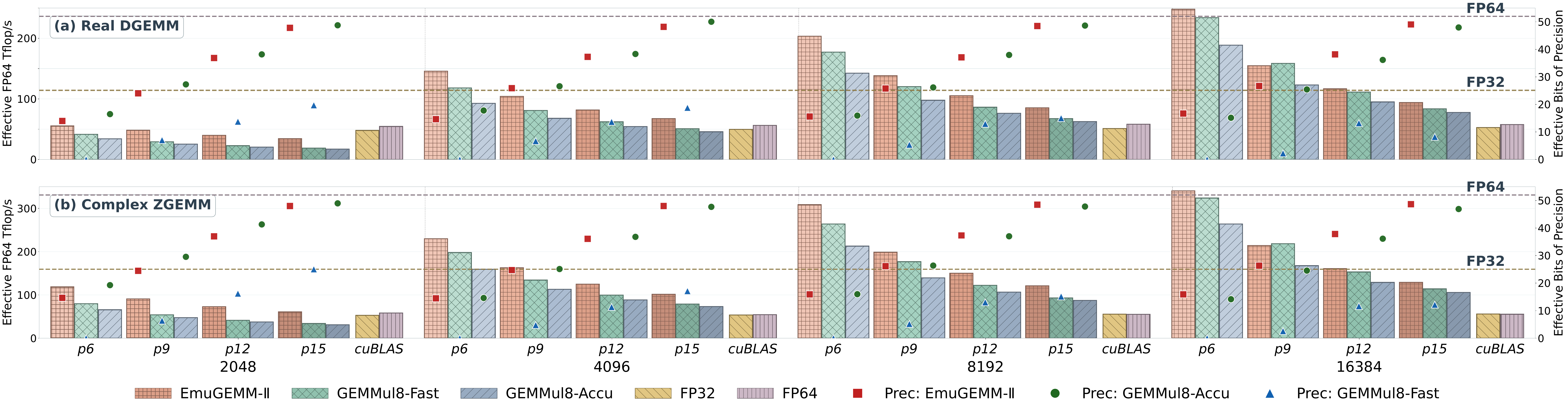}
    \caption{Emulated (a) DGEMM and (b) ZGEMM via 3M on GH200. Bars (left axis) show effective FP64 Tflop/s for $M{=}N{=}K \in \{2048, 4096, 8192, 16384\}$; Markers (right axis) indicate effective bits of precision. EmuGEMM-II is compared to GEMMul8-Fast/Accu at matched $p \in \{6,9,12,15\}$, with cuBLAS FP32/FP64 as reference.}
      \label{fig:dgemm-comparison}
      \vspace{-1em}
\end{figure*}

\begin{figure}[t]
  \centering
  \includegraphics[width=\columnwidth]{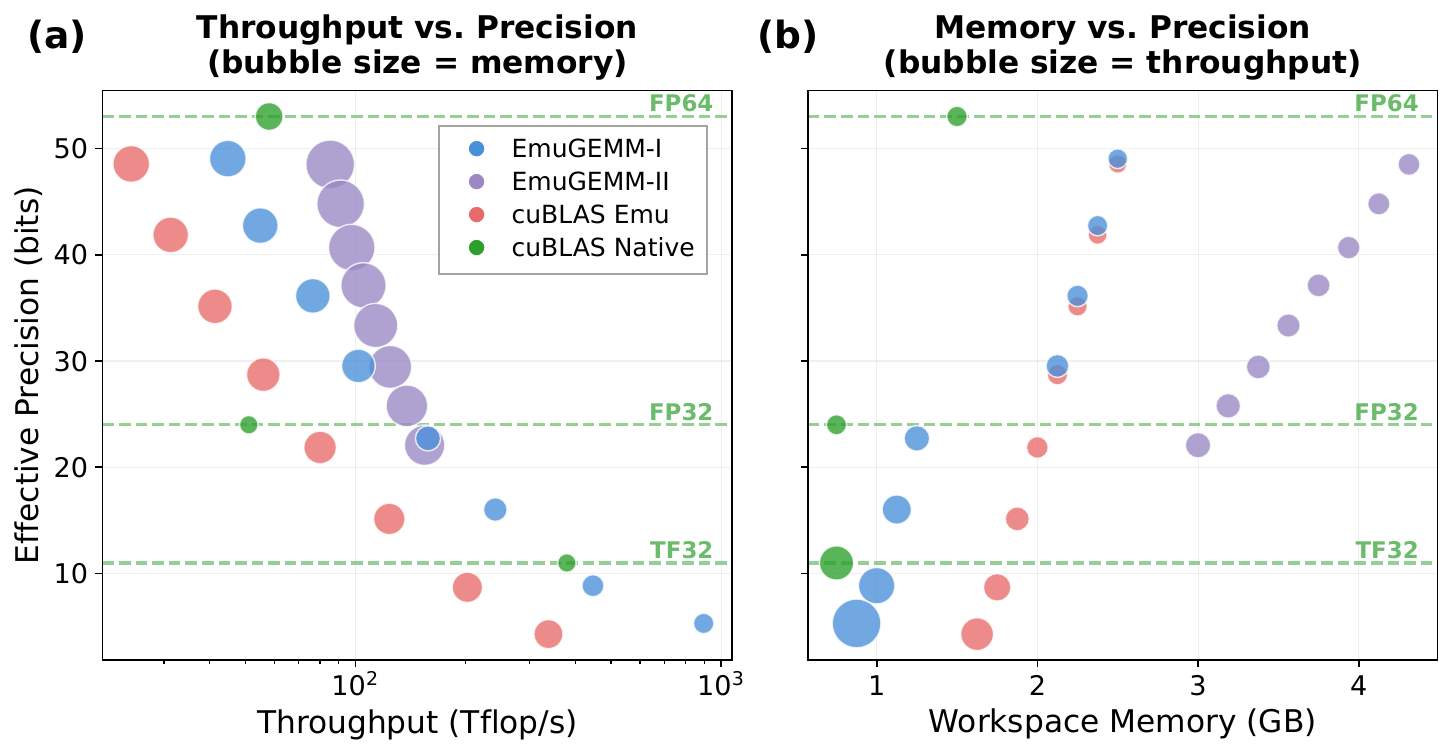}
  \caption{Throughput--precision--memory tradeoff at $M\!=\!N\!=\!K\!=\!8192$ on NVIDIA GH200.}
  \label{fig:tradeoff-2d}
  \vspace{-1.6em}
\end{figure}

\subsection{Comparison with State-of-the-Art Scheme~II}

GEMMul8~\cite{uchino-2025b, uchino-2025c} is the state-of-the-art reference implementation of Ozaki Scheme~II for real and complex GEMM, offering an accuracy-first (GEMMul8-Accu) and a speed-first (GEMMul8-Fast) variant. Figure~\ref{fig:dgemm-comparison} compares EmuGEMM-II against both variants on the GH200 for real DGEMM (sub-plot~(a)) and complex ZGEMM via 3M (sub-plot~(b)). At matched $(p, N)$, EmuGEMM-II is faster than both GEMMul8 variants in nearly all configurations, while its accuracy comes within a few bits of GEMMul8-Accu. The residual trade-off between EmuGEMM-II and GEMMul8-Accu is negligible in practice.

\subsection{Precision-Throughput-Memory Trade-off}

To summarize the trade-offs across all methods, Fig.~\ref{fig:tradeoff-2d} projects cuBLAS native, cuBLAS emulation, EmuGEMM-I, and EmuGEMM-II onto a precision--throughput (sub-plot~(a)) and precision--memory (sub-plot~(b)) plane, at the representative size $M{=}N{=}K{=}8{,}192$ on the GH200.

First, it is observed that EmuGEMM-I and EmuGEMM-II together Pareto-dominate cuBLAS emulation across the entire precision range and outperform cuBLAS native at every operating point above TF32. The crossover between the two schemes lies near FP32 precision: Below FP32, EmuGEMM-I delivers higher throughput; Above FP32, EmuGEMM-II takes over. Together they offer a continuous precision--throughput trade-off from TF32-level to full FP64, whereas cuBLAS native provides only three discrete operating points.

Second, the throughput advantage at high precision comes with a workspace cost. Both schemes allocate $p$ INT8 residue matrices per operand, so operand-side workspace grows linearly with $p$ in both cases. Scheme~II additionally materializes $p$ per-modulus INT8 output residues for CRT reconstruction, pushing its total workspace above Scheme~I at matched $p$. Memory-constrained workloads are, therefore, better suited to Scheme~I, when the precision target permits.

\subsection{Limitations}
\label{sec:discussion-limits}

Our evaluation establishes EmuGEMM's benefits within a defined scope, which we now make explicit.
First, our study is done on limited random datasets, as defined by Eq.~\eqref{eq:test-matrix}.
We deliberately selected a relatively high conditioning parameter to avoid presenting the Ozaki schemes under a best-case scenario.
Still, this analysis is inadequate, but there is extensive prior work on the theoretical and practical accuracy of both Ozaki schemes~\cite{ozaki-1,ozaki-2,uchino-2025a,uchino-2025b,uchino-2025c}.
Our work optimizes the data movement only in the part implemented with INT8 arithmetic, which is exact and avoids floating-point non-associativity.
Therefore, results from existing studies directly apply to our implementations. We do not repeat them here due to space constraints.

Second, we present results only for square matrices of sizes 2,048-16,384.
We have run benchmarks on smaller matrices (512-1,024), which we omit here for brevity, as they add little beyond prior work.
In general, Ozaki-scheme implementations underperform high-precision TCs on small matrices because preprocessing overhead dominates.
Although we observe speedups with our approach over other Ozaki-scheme implementations, they are not enough to surpass the native FP64 units, which remain preferred for small matrices when available.
We also do not test non-square or irregularly-sized matrices.
For example, tall, skinny matrices tend to have lower arithmetic intensity.
However, as our approach is based on standard, highly efficient GEMM implementations, as discussed in Section~\ref{sec:back-gemm}, we do not expect our performance to be more heavily penalized than other approaches.
These evaluation choices are not interface restrictions: EmuGEMM supports arbitrary matrix dimensions, and a batched implementation is possible through a single interleaved tensor per batch entry.
In general, techniques improving GEMM performance on irregular, tall-skinny, or small matrices are orthogonal to our data-movement optimizations and can be combined with them.

Third, we evaluate isolated GEMM calls, whereas applications may execute chains of dependent GEMMs.
As decomposition occurs per operand, an operand shared across several products (e.g., panel updates or repeated weights) is sliced once, amortizing its decomposition cost, while a result feeding a later operation is reconstructed and then re-decomposed, as in cuBLAS emulation and GEMMul8.
Fusing this re-slicing step is left for future work.

Fourth, we acknowledge that our implementations are currently only Hopper- and Blackwell-compatible.
This is in large part due to the use of DeepGEMM, which currently supports only those microarchitectures.
Therefore, further development is needed to evaluate whether speedups similar to those observed in this work can be achieved on other microarchitectures.

\section{Related Work}
\label{sec:related}

Use of lower-precision TCs to accelerate higher-precision GEMM operations is a fast-moving research field.
Early studies characterized the performance and numerical behavior of TCs~\cite{markidis-2018, fasi-2021}, which were then exploited for mixed-precision iterative refinement~\cite{haidar-2018}, error-corrected single-precision GEMM~\cite{ootomo-2022}, the first TC-based Ozaki implementation~\cite{mukunoki-2020}, and exascale mixed-precision benchmarking~\cite{kudo-2020}.
The Ozaki scheme has also been leveraged to accelerate quantum chemistry calculations~\cite{dawson-2024}.
In terms of available Ozaki Scheme~I libraries, the original ozIMMU implementation~\cite{ootomo-2024} and the state-of-the-art one from NVIDIA in cuBLAS~\cite{schwarz2026guaranteed} should be mentioned.
The latter provides several improvements~\cite{schwarz2026guaranteed} over the reference scheme.
First, cuBLAS uses a signed leading slice (SINT8) with unsigned remaining slices (UINT8), gaining one bit of precision per slice over the original signed-everywhere formulation~\cite{ootomo-2024}; our implementation adopts the same scheme.
Second, cuBLAS provides automatic dynamic precision (ADP), which estimates at runtime the slice count required for FP64-level accuracy, at roughly 10\% runtime overhead.
Our implementation lacks such ADP features, which are orthogonal to our proposed data-movement optimizations.
Thus, it is only a matter of engineering effort to integrate them or, inversely, introduce our fused formulation into cuBLAS.

The reference Ozaki Scheme~II implementation can be found in the GEMMul8 library for both real~\cite{uchino-2025b} and complex~\cite{uchino-2025c} floating-point arithmetic.
Our approach directly improves upon these implementations through our fused kernel design, detailed in Section~\ref{sec:emu-ii}.

The Ozaki schemes, as presented so far, use integer arithmetic. However, extensions to non-integer arithmetic are possible: Mukunoki~\cite{mukunoki2025dgemmfp64arithmetic} provides an FP8-based formulation, observing reduced INT8 TC throughput in the NVIDIA B300 GPU, which is different from the B200 GPU discussed in prior sections, compared to emerging low-precision floating-point units.
Although our current implementations do not yet make use of such operations, the proposed data-movement optimizations apply to these extended schemes, and could be included in possible future work. The interleaved layout is indeed slice-format-agnostic (Section~\ref{sec:emu-i-layout}) and the FP8 MMA instructions share the $t_K{=}32$ tile shape of their INT8 counterparts.

\section{Conclusions}
\label{sec:conclusions}

We presented EmuGEMM, a high-performance library for floating-point emulated GEMM.
Our approach focuses on the INT8 formulations of both Ozaki Schemes~I and II and optimizes their data movement via kernel fusion.
The reported benchmarks show significant speedups over the state of the art (cuBLAS emulation, GEMMul8) and over native GEMM performance at comparable accuracy, even on the GH200, which has a very high FP64 tensor-core throughput.
These developments, on top of impactful prior work, may enable even more HPC applications to leverage lower-precision arithmetic on current and future GPU micro-architectures.
As a drop-in GEMM backend, EmuGEMM makes ZGEMM-dominated applications such as \textit{ab initio} quantum transport~\cite{vetsch2025ab} simulations natural integration targets: A separate study~\cite{maeder2026optimizing} demonstrates EmuGEMM's end-to-end benefits in such a case after analyzing the numerical characteristics of its input matrices, achieving up to 51\% higher throughput at maintained accuracy on 40\% fewer HPC resources than its FP64 reference.
Future work could explore how the proposed optimizations can be applied to alternative formulations (e.g., using lower-precision floating-point units),  other GPUs, or even other architectures, for example, CPUs and emerging HPC accelerators.

\section*{Acknowledgment}

This work was supported by the Swiss National Science Foundation (SNSF) under grant $\mathrm{n^\circ}$ 209358 (QuaTrEx), and by the Platform for Advanced Scientific Computing (PASC) in Switzerland (BoostQT). We acknowledge the scientific support and HPC resources from CSCS under projects c33, lp16, and lp82.
The author used Anthropic's Claude as a writing assistant for rephrasing and LaTeX formatting of author-written drafts. All technical content, results, and scientific claims are the authors' own, and all LLM-assisted text was reviewed and verified by the authors.

\bibliographystyle{IEEEtran}
\bibliography{refs}

@article{ozaki-1,
author = {Ozaki, Katsuhisa and Ogita, Takeshi and Oishi, Shin'Ichi and Rump, Siegfried M.},
title = {Error-free transformations of matrix multiplication by using fast routines of matrix multiplication and its applications},
year = {2012},
issue_date = {January   2012},
publisher = {Springer-Verlag},
address = {Berlin, Heidelberg},
volume = {59},
number = {1},
issn = {1017-1398},
url = {https://doi.org/10.1007/s11075-011-9478-1},
doi = {10.1007/s11075-011-9478-1},
journal = {Numer. Algorithms},
month = jan,
pages = {95--118},
numpages = {24}
}

@article{ozaki-2,
      title={Ozaki {Scheme II}: A {GEMM}-oriented emulation of floating-point matrix multiplication using an integer modular technique}, 
      author={Katsuhisa Ozaki and Yuki Uchino and Toshiyuki Imamura},
      journal={The International Journal of High Performance Computing Applications},
      year={2026},
      note={{OnlineFirst}},
      doi={10.1177/10943420261467787},
      url={https://doi.org/10.1177/10943420261467787},
}

@article{ootomo-2024,
    author = {Hiroyuki Ootomo and Katsuhisa Ozaki and Rio Yokota},
    title = {{DGEMM} on integer matrix multiplication unit},
    journal = {The International Journal of High Performance Computing Applications},
    year = {2024},
    volume={38},
    number={4},
    pages={297--313},
    doi = {10.1177/10943420241239588},
    URL = {https://doi.org/10.1177/10943420241239588},
    publisher={SAGE Publications},
}

@article{uchino-2025a,
   title={Performance enhancement of the {Ozaki Scheme} on integer matrix multiplication unit},
   volume={39},
   ISSN={1741-2846},
   url={https://doi.org/10.1177/10943420241313064},
   DOI={10.1177/10943420241313064},
   number={3},
   journal={The International Journal of High Performance Computing Applications},
   publisher={SAGE Publications},
   author={Uchino, Yuki and Ozaki, Katsuhisa and Imamura, Toshiyuki},
   year={2025},
   month=jan,
   pages={462--476}
}

@inproceedings{uchino-2025b,
author = {Uchino, Yuki and Ozaki, Katsuhisa and Imamura, Toshiyuki},
title = {High-Performance and Power-Efficient Emulation of Matrix Multiplication using {INT8} Matrix Engines},
year = {2025},
isbn = {9798400718717},
publisher = {Association for Computing Machinery},
address = {New York, NY, USA},
url = {https://doi.org/10.1145/3731599.3767539},
doi = {10.1145/3731599.3767539},
booktitle = {Proceedings of the SC '25 Workshops of the International Conference for High Performance Computing, Networking, Storage and Analysis},
pages = {1824–1831},
numpages = {8},
location = {
},
series = {SC Workshops '25}
}

@inproceedings{uchino-2025c,
      title={Emulation of Complex Matrix Multiplication based on the {Chinese Remainder Theorem}},
      author={Yuki Uchino and Qianxiang Ma and Toshiyuki Imamura and Katsuhisa Ozaki and Patrick Lars Gutsche},
      booktitle={ISC High Performance 2026 Research Paper Proceedings (41st International Conference)},
      address={Hamburg, Germany},
      publisher={IEEE},
      year={2026},
      pages={1--12},
      doi={10.23919/ISC.2026.11520500},
      url={https://doi.org/10.23919/ISC.2026.11520500}, 
}

@misc{nvidia_hopper,
  title={{NVIDIA H100 Tensor Core GPU} Architecture Whitepaper},
  author={{NVIDIA Corporation}},
  year={2022},
  howpublished={\url{https://resources.nvidia.com/en-us-hopper-architecture}}
}

@misc{nv_tensor_core,
  title={{NVIDIA} Tensor Cores},
  author={{NVIDIA Corporation}},
  year={2025},
  howpublished={\url{https://www.nvidia.com/en-us/data-center/tensor-cores/}}
}

@misc{amd_cdna,
  title   = {{AMD CDNA Architecture}},
  author  = {{Advanced Micro Devices}},
  year    = {2026},
  howpublished={\url{https://www.amd.com/en/technologies/cdna.html}}
}

@article{higham2019simulating,
author = {Higham, Nicholas J. and Pranesh, Srikara},
title = {Simulating Low Precision Floating-Point Arithmetic},
journal = {SIAM Journal on Scientific Computing},
volume = {41},
number = {5},
pages = {C585-C602},
year = {2019},
doi = {10.1137/19M1251308},
URL = {https://doi.org/10.1137/19M1251308},
}

@article{blanchard2020mixed,
author = {Blanchard, Pierre and Higham, Nicholas J. and Lopez, Florent and Mary, Theo and Pranesh, Srikara},
title = {Mixed Precision Block Fused Multiply-Add: Error Analysis and Application to GPU Tensor Cores},
journal = {SIAM Journal on Scientific Computing},
volume = {42},
number = {3},
pages = {C124-C141},
year = {2020},
doi = {10.1137/19M1289546},
URL = {https://doi.org/10.1137/19M1289546},
}

@misc{vastai,
  author       = {{Vast.ai, Inc.}},
  title        = {{Vast.ai} {GPU} Compute Marketplace},
  howpublished = {\url{https://vast.ai}},
  year         = {2026},
  note         = {Accessed: 2026-04-05}
}

@inproceedings{schwarz2026guaranteed,
author = {Schwarz, Angelika and Anders, Anton and Brower, Cole and Bayraktar, Harun and Gunnels, John and Clark, Kate and Xu, RuQing G. and Rodriguez, Samuel and Cayrols, Sebastien and Tabaszewski, Pawel and Podlozhnyuk, Victor},
title = {Guaranteed {DGEMM} Accuracy While Using Reduced Precision Tensor Cores Through Extensions of the {Ozaki} Scheme},
year = {2026},
isbn = {9798400720673},
publisher = {Association for Computing Machinery},
address = {New York, NY, USA},
url = {https://doi.org/10.1145/3773656.3773670},
doi = {10.1145/3773656.3773670},
booktitle = {Proceedings of the Supercomputing Asia and International Conference on High Performance Computing in Asia Pacific Region},
pages = {91--101},
numpages = {11},
location = {
},
series = {SCA/HPCAsia '26}
}

@inproceedings{dao2022flashattention,
 author = {Dao, Tri and Fu, Dan and Ermon, Stefano and Rudra, Atri and R\'{e}, Christopher},
 booktitle = {Advances in Neural Information Processing Systems},
 editor = {S. Koyejo and S. Mohamed and A. Agarwal and D. Belgrave and K. Cho and A. Oh},
 pages = {16344--16359},
 publisher = {Curran Associates, Inc.},
 title = {{FlashAttention}: Fast and Memory-Efficient Exact Attention with IO-Awareness},
 url = {https://proceedings.neurips.cc/paper_files/paper/2022/file/67d57c32e20fd0a7a302cb81d36e40d5-Paper-Conference.pdf},
 volume = {35},
 year = {2022}
}

@inproceedings{vetsch2025ab,
author = {Vetsch, Nicolas and Maeder, Alexander and Maillou, Vincent and Winka, Anders and Cao, Jiang and Kwasniewski, Grzegorz and Deuschle, Leonard and Hoefler, Torsten and Ziogas, Alexandros Nikolaos and Luisier, Mathieu},
title = {Ab-initio Quantum Transport with the {GW} Approximation, 42,240 Atoms, and Sustained Exascale Performance},
year = {2025},
isbn = {9798400714665},
publisher = {Association for Computing Machinery},
address = {New York, NY, USA},
url = {https://doi.org/10.1145/3712285.3771784},
doi = {10.1145/3712285.3771784},
booktitle = {Proceedings of the International Conference for High Performance Computing, Networking, Storage and Analysis},
pages = {1--13},
numpages = {13},
location = {
},
series = {SC '25}
}

@misc{deepgemm,
      title={{DeepGEMM}: clean and efficient {BLAS} kernel library on {GPU}}, 
      author={Chenggang Zhao and Zhean Xu and Liang Zhao and Jiashi Li and Chenhao Xu and Anyi Xu and Shengyu Liu and Kexing Zhou and Kuai Yu},
      year={2025},
      publisher = {GitHub},
      howpublished = {\url{https://github.com/deepseek-ai/DeepGEMM}},
}

@misc{cecka2026cutelayoutrepresentationalgebra,
      title={{CuTe} Layout Representation and Algebra}, 
      author={Cris Cecka},
      year={2026},
      eprint={2603.02298},
      archivePrefix={arXiv},
      primaryClass={cs.MS},
      url={https://doi.org/10.48550/arXiv.2603.02298}, 
}

@software{Thakkar_CUTLASS_2023,
author = {Thakkar, Vijay and Ramani, Pradeep and Cecka, Cris and Shivam, Aniket and Lu, Honghao and Yan, Ethan and Kosaian, Jack and Hoemmen, Mark and Wu, Haicheng and Kerr, Andrew and Nicely, Matt and Merrill, Duane and Blasig, Dustyn and Atluri, Aditya and Qiao, Fengqi and Majcher, Piotr and Springer, Paul and Hohnerbach, Markus and Wang, Jin and Gupta, Manish},
license = {BSD-3-Clause},
month = jan,
title = {{CUTLASS}},
url = {https://github.com/NVIDIA/cutlass},
version = {3.0.0},
year = {2023}
}

@misc{gemmul8,
  title={{GEMMul8} ({GEMMulate}): {GEMM} emulation and its extension to {BLAS}-like matrix operations using {INT8}/{FP8} matrix engines based on the {Ozaki Scheme II}},
  author={{RIKEN R-CCS}},
  year={2025},
  howpublished={\url{https://github.com/RIKEN-RCCS/GEMMul8}}
}

@inproceedings{mukunoki2025dgemmfp64arithmetic,
      title={{DGEMM} using {FP64} Arithmetic Emulation and {FP8} Tensor Cores with {Ozaki} Scheme}, 
      author={Daichi Mukunoki},
      booktitle={Proceedings of the Supercomputing Asia and International Conference on High Performance Computing in Asia Pacific Region Workshops},
      series={SCA/HPCAsiaWS '26},
      address={Osaka, Japan},
      publisher={Association for Computing Machinery},
      year={2026},
      pages={303--311},
      doi={10.1145/3784828.3785017},
      url={https://doi.org/10.1145/3784828.3785017}
}

@inproceedings{mukunoki-2020,
      title={{DGEMM} Using Tensor Cores, and Its Accurate and Reproducible Versions},
      author={Daichi Mukunoki and Katsuhisa Ozaki and Takeshi Ogita and Toshiyuki Imamura},
      booktitle={High Performance Computing: 35th International Conference, {ISC} High Performance 2020},
      series={Lecture Notes in Computer Science},
      volume={12151},
      pages={230--248},
      publisher={Springer},
      address={Cham},
      year={2020},
      doi={10.1007/978-3-030-50743-5_12},
      url={https://doi.org/10.1007/978-3-030-50743-5_12},
}

@article{ootomo-2022,
      title={Recovering single precision accuracy from {Tensor Cores} while surpassing the {FP32} theoretical peak performance},
      author={Hiroyuki Ootomo and Rio Yokota},
      journal={The International Journal of High Performance Computing Applications},
      volume={36},
      number={4},
      pages={475--491},
      year={2022},
      doi={10.1177/10943420221090256},
      url={https://doi.org/10.1177/10943420221090256},
}

@article{fasi-2021,
      title={Numerical behavior of {NVIDIA} tensor cores},
      author={Massimiliano Fasi and Nicholas J. Higham and Mantas Mikaitis and Srikara Pranesh},
      journal={PeerJ Computer Science},
      volume={7},
      pages={e330},
      year={2021},
      doi={10.7717/peerj-cs.330},
      url={https://doi.org/10.7717/peerj-cs.330},
}

@inproceedings{markidis-2018,
      title={{NVIDIA} Tensor Core Programmability, Performance \& Precision},
      author={Stefano Markidis and Steven Wei Der Chien and Erwin Laure and Ivy Bo Peng and Jeffrey S. Vetter},
      booktitle={2018 IEEE International Parallel and Distributed Processing Symposium Workshops (IPDPSW)},
      pages={522--531},
      publisher={IEEE},
      year={2018},
      doi={10.1109/IPDPSW.2018.00091},
      url={https://doi.org/10.1109/IPDPSW.2018.00091},
}

@inproceedings{haidar-2018,
      title={Harnessing {GPU} Tensor Cores for Fast {FP16} Arithmetic to Speed up Mixed-Precision Iterative Refinement Solvers},
      author={Azzam Haidar and Stanimire Tomov and Jack Dongarra and Nicholas J. Higham},
      booktitle={SC18: International Conference for High Performance Computing, Networking, Storage and Analysis},
      pages={603--613},
      publisher={IEEE},
      year={2018},
      doi={10.1109/SC.2018.00050},
      url={https://doi.org/10.1109/SC.2018.00050},
}

@article{abdelfattah-2021,
      title={A survey of numerical linear algebra methods utilizing mixed-precision arithmetic},
      author={Ahmad Abdelfattah and Hartwig Anzt and Erik G. Boman and Erin Carson and Terry Cojean and Jack Dongarra and Alyson Fox and Mark Gates and Nicholas J. Higham and Xiaoye S. Li and Jennifer Loe and Piotr Luszczek and Srikara Pranesh and Siva Rajamanickam and Tobias Ribizel and Barry F. Smith and Kasia Swirydowicz and Stephen Thomas and Stanimire Tomov and Yaohung M. Tsai and Ulrike Meier Yang},
      journal={The International Journal of High Performance Computing Applications},
      volume={35},
      number={4},
      pages={344--369},
      year={2021},
      doi={10.1177/10943420211003313},
      url={https://doi.org/10.1177/10943420211003313},
}

@article{higham-1992,
      title={Stability of a Method for Multiplying Complex Matrices with Three Real Matrix Multiplications},
      author={Nicholas J. Higham},
      journal={SIAM Journal on Matrix Analysis and Applications},
      volume={13},
      number={3},
      pages={681--687},
      year={1992},
      doi={10.1137/0613043},
      url={https://doi.org/10.1137/0613043},
}

@article{rump-2008,
      title={Accurate Floating-Point Summation Part {I}: Faithful Rounding},
      author={Siegfried M. Rump and Takeshi Ogita and Shin'ichi Oishi},
      journal={SIAM Journal on Scientific Computing},
      volume={31},
      number={1},
      pages={189--224},
      year={2008},
      doi={10.1137/050645671},
      url={https://doi.org/10.1137/050645671},
}

@inproceedings{kudo-2020,
      title={Implementation and Numerical Techniques for One {EFlop/s} {HPL-AI} Benchmark on {Fugaku}},
      author={Shuhei Kudo and Keigo Nitadori and Takuya Ina and Toshiyuki Imamura},
      booktitle={2020 IEEE/ACM 11th Workshop on Latest Advances in Scalable Algorithms for Large-Scale Systems (ScalA)},
      pages={69--76},
      publisher={IEEE},
      year={2020},
      doi={10.1109/ScalA51936.2020.00014},
      url={https://doi.org/10.1109/ScalA51936.2020.00014},
}

@article{dawson-2024,
      title={Reducing Numerical Precision Requirements in Quantum Chemistry Calculations},
      author={William Dawson and Katsuhisa Ozaki and Jens Domke and Takahito Nakajima},
      journal={Journal of Chemical Theory and Computation},
      volume={20},
      number={24},
      pages={10826--10837},
      year={2024},
      doi={10.1021/acs.jctc.4c00938},
      url={https://doi.org/10.1021/acs.jctc.4c00938},
}

@article{vanzee-2015,
      title={{BLIS}: A Framework for Rapidly Instantiating {BLAS} Functionality},
      author={Field G. {Van Zee} and Robert A. {van de Geijn}},
      journal={ACM Transactions on Mathematical Software},
      volume={41},
      number={3},
      pages={1--33},
      year={2015},
      doi={10.1145/2764454},
      url={https://doi.org/10.1145/2764454},
}

@inproceedings{jouppi-2017,
      title={In-Datacenter Performance Analysis of a Tensor Processing Unit},
      author={Norman P. Jouppi and Cliff Young and Nishant Patil and David Patterson and others},
      booktitle={Proceedings of the 44th Annual International Symposium on Computer Architecture},
      series={ISCA '17},
      publisher={Association for Computing Machinery},
      pages={1--12},
      year={2017},
      doi={10.1145/3079856.3080246},
      url={https://doi.org/10.1145/3079856.3080246},
}

@misc{maeder2026optimizing,
      title={Optimizing Semiconductor Device Simulations through Low-Precision Arithmetic},
      author={Alexander Maeder and Denghui Lu and Nicolas Vetsch and Vincent Maillou and Anders Winka and Jiang Cao and Mauro Dossena and Alexandros Nikolaos Ziogas and Mathieu Luisier},
      year={2026},
      eprint={2606.25595},
      archivePrefix={arXiv},
      primaryClass={cs.CE},
      url={https://doi.org/10.48550/arXiv.2606.25595},
}

\end{document}